\documentclass[final,5p,times,twocolumn]{elsarticle}
\journal{Nuclear Physics B}

\usepackage[utf8]{inputenc}
\usepackage{graphicx}
\usepackage{dcolumn}
\usepackage{bm}
\usepackage{amsmath}
\usepackage{amsfonts}
\usepackage{slashed}
\usepackage{hyperref}
\usepackage{multirow}
\usepackage{soul}
\usepackage{mathtools}
\usepackage{float}
\usepackage{afterpage}
\usepackage{color}

\usepackage{amsmath,amssymb}
\usepackage{graphicx}
\usepackage{bm}
\usepackage{comment} 
\usepackage{subfigure}
\usepackage{array}
\usepackage{multirow} 

\DeclareMathOperator{\Tr}{Tr}

\newcommand{\nn}{\nonumber}

\newcommand{\Erho}{ \left[755(2)(1)(^{20}_{02})-\frac{i}{2}\,129(3)(1)(^{7}_{1})\right]~{\rm MeV}}

\hypersetup{colorlinks, linkcolor = [rgb]{0,0.0,0.75}, citecolor = [rgb]{0,0.0,0.75}, urlcolor = [rgb]{0,0.0,0.75}}

\begin{document}

\begin{frontmatter}
 
\title{Connecting physical resonant amplitudes and lattice QCD}

\author[drb1,drb2]{Daniel R. Bolton\corref{cor1}}
\ead{daniel.bolton@colorado.edu}
\address[drb1]{Department of Physics, University of Colorado, Boulder, CO 80309, USA}
\address[drb2]{Department of Physics, Baylor University, Waco, TX 76798, USA}
\cortext[cor1]{Corresponding author}
\author[rab]{Ra\'ul A. Brice\~no}
\ead{rbriceno@jlab.org}
\address[rab]{Thomas Jefferson National Accelerator Facility, 12000 Jefferson Avenue, Newport News, VA 23606, USA}
\author[djw]{David J. Wilson}
\ead{djwilson@jlab.org}
\address[djw]{Department of Physics, Old Dominion University, Norfolk, VA 23529, USA}

\begin{abstract}
 We present a determination of the isovector, $P$-wave $\pi\pi$ scattering phase shift obtained by extrapolating recent lattice QCD results from the Hadron Spectrum Collaboration using $m_\pi =236$ MeV. The finite volume spectra are described using extensions of L\"uscher's method to determine the infinite volume Unitarized Chiral Perturbation Theory scattering amplitude. We exploit the pion mass dependence of this effective theory to obtain the scattering amplitude at $m_\pi= 140$ MeV. The scattering phase shift is found to agree with experiment up to center of mass energies of 1.2~GeV. The analytic continuation of the scattering amplitude to the complex plane yields a $\rho$-resonance pole at $E_\rho= \left[755(2)(1)(^{20}_{02})-\frac{i}{2}\,129(3)(1)(^{7}_{1})\right]~{\rm MeV}$. The techniques presented illustrate a possible pathway towards connecting lattice QCD observables of few-body, strongly interacting systems to experimentally accessible quantities. 
\end{abstract}
 
\begin{keyword}
Lattice QCD \sep Chiral Perturbation Theory \sep Pion elastic scattering \sep \href{http://arxiv.org/abs/1507.07928}{arXiv:1507.07928}
\end{keyword}

\end{frontmatter}


The spectrum of hadronic resonances has long served as a window into the non-perturbative nature of Quantum Chromodynamics (QCD), the fundamental theory of the strong force. Hadronic resonances are color-singlet combinations of the fundamental degrees of freedom of QCD (quarks, anti-quarks, and gluons). They are observed as unstable resonant enhancements in the scattering of QCD stable hadrons, such as the pion. A simple example of a hadronic resonance is the $\rho$ that occurs in $\pi\pi$ scattering. The non-perturbative nature of QCD makes direct determination of the properties of hadronic resonances a challenging task.

Presently, the only means to study properties of low-energy hadronic states in a systematically improvable way is to perform a non-perturbative numerical evaluation of the QCD path-integral, by statistically sampling the gauge fields in a discretized finite volume to obtain correlation functions. This program is known as lattice QCD. The last decade has witnessed a tremendous advance in the ability of the lattice QCD community to connect experimental phenomena directly to the standard model of particle physics. It is not unreasonable to expect that in the upcoming decade most ``\emph{simple}" observables, such as masses, decay constants and elastic form factors of low-lying QCD stable particles, will be computed using physical values of the quark masses and QCD+QED gauge configurations (see Refs.~\cite{Borsanyi:2014jba, Borsanyi:2013lga, Aoki:2012st} for recent progress in this direction). 

For hadronic resonances, and in general systems involving two or more stable hadrons, the challenges are far greater and further technological and formal developments are needed (see Refs.~\cite{Briceno:2014tqa, Briceno:2014pka, Yamazaki:2015nka, Prelovsek:2014zga} for recent reviews on the topic). In order to kinematically suppress multiparticle channels, many excited state calculations are performed using unphysically massive light quarks. Thus, it is desirable to devise a scheme for performing a controlled extrapolation to the physical mass.

As a step towards developing such a program, we present the first extrapolation of a resonant scattering amplitude obtained from lattice QCD. Specifically, we analyze isovector, $P$-wave $\pi\pi$ spectra in the elastic scattering region that have been determined by the \emph{Hadron Spectrum Collaboration} using dynamical quark masses corresponding to $m_\pi= 236$~MeV~\cite{Wilson:2015dqa}.
 
Lattice QCD uses a discrete and finite spacetime. Discretization provides a natural high energy regulator for QCD and if a fine enough spacing is used this introduces negligibly small effects in the spectrum. Working in a finite, periodic volume transforms the continuum of infinite volume scattering states into a discrete spectrum of states. The non-perturbative \emph{mapping} between finite and infinite volume observables was first derived in Refs.~\cite{Luscher:1986pf, Luscher:1990ux} and is commonly referred to as the ``L\"uscher method''.

The mappings between finite and infinite volume amplitudes cannot be one-to-one due to two important facts. First, the reduction of rotational symmetry from a continuous group to a discrete group (e.g., cubic) assures mixing between different partial waves. Second, having lost the notion of asymptotic states, finite volume states will necessarily be an admixture of different hadronic states with the same quantum numbers (e.g., $\pi\pi$ and $K\overline{K}$ in the $I=1$ channel). Many theoretical advances have guided the field. For example, several references have discussed the feasibility of studying coupled-channel scattering in a finite volume~\cite{He:2005ey, Briceno:2012yi, Hansen:2012tf,  Briceno:2014oea} (see Refs.~\cite{Dudek:2014qha, Wilson:2014cna} for the first application of this formalism to the study of $\pi K, \eta K$) as well as three-body systems~\cite{Hansen:2014eka, Hansen:2015zga, Briceno:2012rv, Polejaeva:2012ut}. These methods become increasingly cumbersome when applied to highly energetic few-body systems, such as exotic or hybrid resonances~\cite{Dudek:2009qf, Dudek:2011tt, Liu:2012ze}, as well as the phenomenologically interesting charm and bottom decays (e.g., $D\rightarrow \pi\pi/K\overline{K}$ \cite{Hansen:2012tf, Aaij:2011in}), where multiple few-body channels are open.

In this work, we investigate one of the most studied low-lying resonances, the $\rho$~\cite{Wilson:2015dqa, Metivet:2014bga, Dudek:2012xn, Feng:2010es, Lang:2011mn, Pelissier:2012pi, Aoki:2011yj, Aoki:2007rd}. The $\rho$ is an isotriplet with $J^{PC}=1^{--}$, and it decays strongly to $\pi\pi$ nearly 100\% of the time~\cite{Agashe:2014kda}. Its mass, $\sim770$~MeV, lies above the $\pi\pi$ and  $4\pi$ thresholds, and is less than half a width [$\Gamma_\rho\sim145$~MeV] away from the $6\pi$ threshold. The coupling to these channels are experimentally observed to be negligible, which would suggest that the finite volume effects associated with these thresholds are suppressed. Further work is needed to confirm and quantify this suppression.

To circumvent these subtleties, we perform an extrapolation to the physical point of the $\pi\pi$ scattering phase shift computed at $m_\pi= 236$~MeV~\cite{Wilson:2015dqa}. At these quark masses, the $4\pi$, $6\pi$ and $K\overline{K}$ thresholds lie well above the $\rho$ resonance and can be safely ignored. To perform the extrapolation we use \emph{Unitarized Chiral Perturbation Theory} (U$\chi$PT)~\cite{Oller:1997ng, Dobado:1996ps, Oller:1998hw, GomezNicola:2001as, Pelaez:2006nj}, which we summarize below. The parameters of U$\chi$PT at $m_\pi=236$~MeV are chosen in order to reproduce the lattice QCD spectrum, and once this is done the pion mass is set to its experimental value and a postdiction for the scattering phase shift is obtained. Although superficially the need to extrapolate may seem undesirable, the avoidance of thresholds makes this conjunction of a phenomenological effective field theory with the L\"uscher method a fruitful alternative to a determination of the phase shift at the physical point.

U$\chi$PT was previously advocated in the literature as a tool to determine physical resonances from lattice QCD~\cite{Chen:2012rp, Doring:2012eu, Doring:2011nd, Doring:2011vk, Bernard:2010fp, Nebreda:2011di, Rios:2008zr, Guo:2008nc, Hanhart:2008mx}, and it has been used in the study of the quark-mass dependence of the $\rho$ mass~\cite{Pelaez:2010fj}~\footnote{It also has been used to determine the low-energy coefficients (LECs) for heavy-light systems by studying the quark-mass dependence of the scattering phase shifts of weakly repulsive channels~\cite{Liu:2012zya}.} . Instead of focusing on the pole of the resonant amplitude, which has been the main focus of previous chiral extrapolations, we fit the full resonant amplitude. Given the correlation between the energy- and quark-mass dependence of these amplitudes, we find that this is sufficient to obtain the quark-mass dependence of the amplitude and consequently its pole. 

In ref.~\cite{Wilson:2015dqa}, a total of 22 $\pi\pi$ energy levels are obtained below the $4\pi/K\overline{K}$ thresholds. Also determined are energy levels above these thresholds, and from them the $K\overline{K}$ phase shift and $\pi\pi,K\overline{K}$ inelasticity are obtained using the formalism first presented in~\cite{Briceno:2012yi, Hansen:2012tf}. In this work, we analyze only the states in the elastic region. To relate these to an infinite volume scattering amplitude, $\mathcal M(P)$, we use the generalization of L\"uscher's formalism for two degenerate scalar particles in moving frames~\cite{Luscher:1986pf, Luscher:1990ux, Rummukainen:1995vs, Kim:2005gf, Christ:2005gi}
\begin{equation}
\label{eq:QC}
\det[F^{-1}(P,L) + \mathcal M(P)] = 0\,,
\end{equation}
where $F(P,L)$ is a function that depends on the total four-momentum $P$ and the spatial extent of the cubic volume $L$, and the determinant acts on the space of spherical harmonics (for an exact definition of these quantities see Ref.~\cite{Kim:2005gf}). This expression is exact up to exponentially suppressed corrections that scale as $e^{-m_\pi L}$, which we can safely ignore given that $m_\pi L\approx4.4$ for the lattice used~\cite{Wilson:2015dqa}
~\footnote{A subset of these exponential corrections has been determined for the $\pi\pi$ states with $\ell=0$~\cite{Bedaque:2006yi} and $\ell=1$~\cite{Chen:2012rp} partial waves.}. Because the two particles are degenerate, odd and even partial waves do not couple, even when the system is in flight. Furthermore, in Ref.~\cite{Wilson:2015dqa} it was shown that in the elastic region the $\ell\geq3$ phase shifts are consistent with zero. Therefore, Eq.~\ref{eq:QC} effectively gives a one-to-one relation between the spectrum and the elastic $(\ell,I)=(1,1)$ $\pi\pi$ scattering amplitude. For real values of the relative momentum, $q$, the inverse of the scattering amplitude is related to the scattering phase shift $\delta$ in the standard way~\cite{Kim:2005gf}
\begin{equation}
q\cot\delta_\ell^I=16\pi E_{\pi\pi}^\star{\rm Re}\left[\left({\cal M}^I_\ell\right)^{-1}\right],
\end{equation}
where $E_{\pi\pi}^\star=2\sqrt{q^2+m^2_\pi}$ is the total energy in the center of mass (c.m.) frame.

We use SU(2) U$\chi$PT to obtain the $\pi\pi$ amplitude. Just like standard $\chi$PT~\cite{Weinberg:1966kf, Colangelo:2001df,Ecker:1988te,Gasser:1984gg,Gasser:1983yg,Gasser:1983kx}, U$\chi$PT allows one to evaluate observables analytically in a perturbative expansion defined by $\left({m_\pi}/{4\pi f_\pi}\right)^2$, where $f_\pi=92.2~\rm MeV$~\cite{Agashe:2014kda} is the decay constant of the $\pi$. At each order in the expansion, one can write the scattering amplitude as a function of a finite number of LECs. At leading-order (LO) in the expansion only two LECs appear ($m_0$ and $f_0$). At next-to-leading order (NLO) four other LECs emerge ($\ell^r_{i=1-4}$). See~\ref{sec:ChPT} for the Lagrangian as well as perturbative expressions for the pion mass, decay constant, and the pion-pion scattering amplitude. When performing the fit to the lattice spectrum, we fix $m_0$ such that $m_\pi=236$ MeV. Given that the decay constant has not been determined, $f_0$ is fixed to reproduce the experimental value of $f_\pi$.
~\footnote{For progress towards determining the decay constant of the ground state and excited states of the $\pi$ using these lattices, we point the reader to Ref.~\cite{Mastropas:2014fsa}.} 
The $\ell^r_i$ cannot be directly obtained from the physical values of the mass and decay constant, but can be accessed from the scattering amplitude. For the $\ell=1$ partial wave, only two linear combinations of these are needed to describe the scattering phase shift ($\alpha_1\equiv-2\ell^r_1+\ell^r_2$ and $\alpha_2\equiv\ell^r_4$). As discussed below, we fix these parameters by performing a fit to the lattice spectrum. Although the $\ell^r_i$ are quark-mass independent in principle, by ignoring higher-order corrections the LECs will absorb a mild quark-mass dependence. See Ref.~\cite{Durr:2014oba} for a recent review and discussion in the context of standard $\chi$PT
\footnote{In Ref.~\cite{Pelaez:2010fj} it is argued that these effects might be large for U$\chi$PT and higher order corrections might be needed. In this work we ignored higher order corrections, and these will be incorporated in future studies.}.

The distinguishing feature of U$\chi$PT is its use of a procedure commonly referred to in the literature as the \emph{Inverse Amplitude Method}~\cite{Oller:1997ng, Oller:1998hw, GomezNicola:2001as} to ensure that the scattering amplitude satisfies unitarity.  Effectively, in U$\chi$PT $s$-channel diagrams are summed in a geometric series using perturbation theory to all orders, while $t$- and $u$-channel diagrams are treated perturbatively to a finite order in the expansion described above
\footnote{We point the reader to Ref.~\cite{GomezNicola:2007qj,Pelaez:2010fj} for a rigorous derivation using dispersive techniques~\cite{Ananthanarayan:2000ht, GarciaMartin:2011cn, Danilkin:2011fz}. The authors are not aware of such a derivation for inelastic processes, e.g., $\pi K\to \eta K$~\cite{Dudek:2014qha, Wilson:2014cna}.}. 
This procedure empirically extends the range of applicability of standard $\chi$PT to c.m. energies on the order of 1.2~GeV. Furthermore, unlike standard $\chi$PT, U$\chi$PT has been shown to accurately describe low-lying resonances with a finite number of LECs~\cite{Oller:1997ng, Oller:1998hw, GomezNicola:2001as}, making it a desirable tool for the study of resonances from lattice QCD. By truncating the chiral expansion to NLO, one can write the unitarized scattering amplitude (see~\ref{sec:IAmpMeth} for the derivation),
\begin{align}
\mathcal{M}_{\rm U\chi PT}
=\mathcal{M}_{\rm LO}\frac{1}{\mathcal{M}_{\rm LO}-\mathcal{M}_{\rm NLO}}
\mathcal{M}_{\rm LO},\label{eq:MUChPT}
\end{align}
where $\mathcal{M}_{\rm LO}$ and $\mathcal{M}_{\rm NLO}$ are the LO and NLO $\chi PT$ amplitudes detailed in~\ref{sec:ChPT}.

\begin{figure}[t]
\begin{center}
\hspace*{-.5cm}                                                           
\includegraphics[scale=0.38]{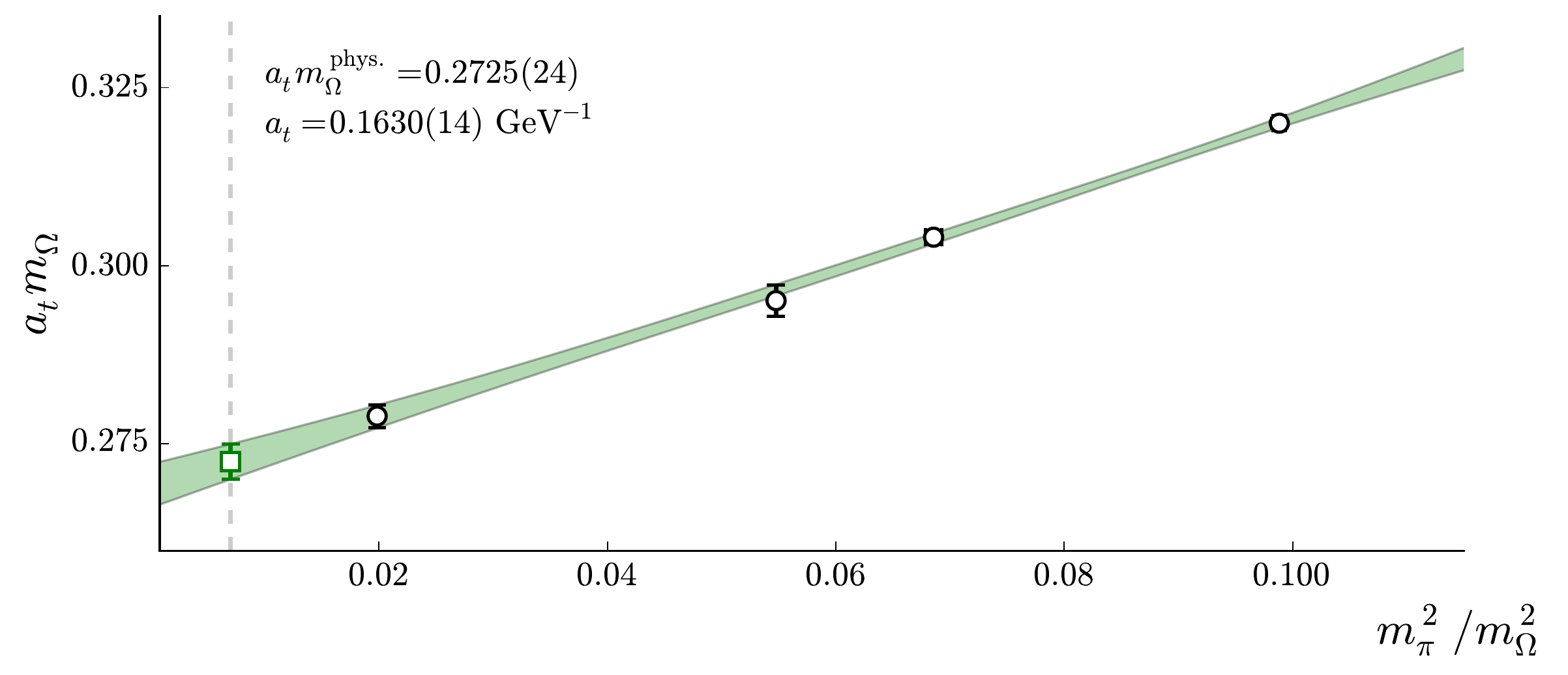}
\caption{Shown is the values of  $a_tm_\Omega$ previously determined for four different values of quark masses (black circles) ~\cite{Lin:2008pr, Wilson:2015dqa}. The green band depicts the fit to these masses using Eq.~\ref{eq:m_omega}. The physical point is denoted by the dashed line. By fixing the resulting value of $a_tm_\Omega$ to $a_tm_\Omega^{\rm phys.}$ we obtain $ a^{[2]}_t=0.1630(14)~{\rm GeV}^{-1}$. }
\label{fig:m_omega}
\end{center}
\end{figure}

To perform a chiral extrapolation we must determine the lattice spacing. We use two definitions of the lattice spacing. First, we use the $\Omega$ baryon mass, which has been determined to be $a_tm_\Omega^{\rm latt.}=0.2789(16)$ at these quark masses~\cite{Wilson:2015dqa}. By setting this equal to $a_tm_\Omega^{\rm phys.}$, where $m_\Omega^{\rm phys.}=1672.45(29)~\rm MeV$ is the mass of physical $\Omega$ baryon, we obtain the lattice spacing $ a_t^{[1]}=0.1668(10)~{\rm GeV}^{-1}$.  Second, as shown in Fig. \ref{fig:m_omega}, we perform an extrapolation to the physical point of the lattice $\Omega$ baryon mass using 
\begin{align}
\label{eq:m_omega}
m_\Omega(m_\pi)=m_{\Omega,0}+\alpha \frac{m_\pi^2}{m_\Omega^2}+\beta \frac{m_\pi^4}{m_\Omega^4}
\end{align}
determined for four different values of $\frac{a_t m_\pi}{a_t m_\Omega}\in[0.14-0.33]$~\cite{Lin:2008pr, Wilson:2015dqa}. We find $ a^{[2]}_t=0.1630(14)~{\rm GeV}^{-1}$ with a $\rm \chi^2/d.o.f.=0.52$. Assuming that $a_t^{[1]}$ should coincide with $a_t^{[2]}$, we perform all fits using both of these lattice spacings and any deviation of the result is incorporated into the systematic error. All central values below are obtained using the mean value of $a_t^{[1]}$. As shown below, this $2\%$ error is the largest source of uncertainty in our final result. It is important to recognize that this systematic error is improvable.

\begin{figure}[t]
\begin{center}
\hspace*{-.5cm}                                                           
\includegraphics[scale=0.38]{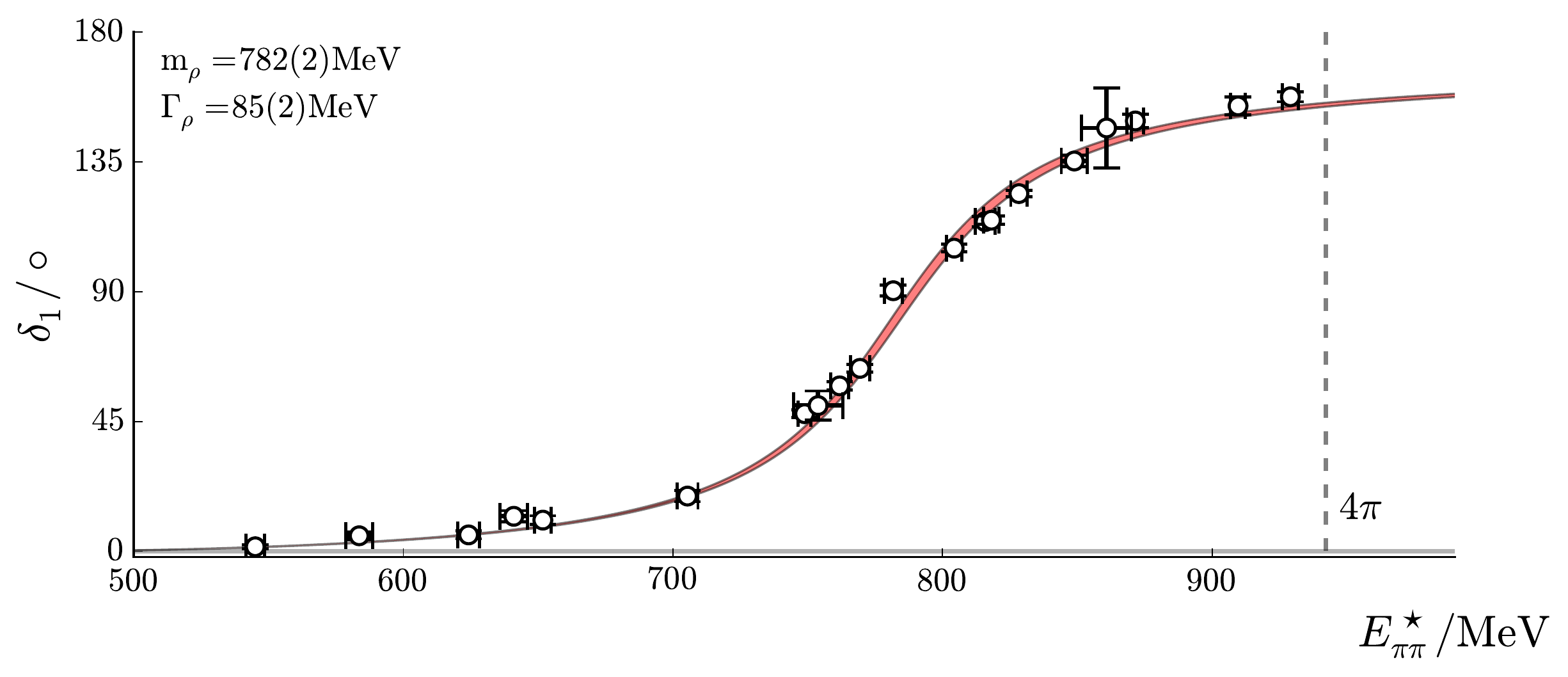}
\caption{Shown is the $I=1$ $\pi\pi$ phase shift obtained from the lattice QCD spectrum determined at $m_\pi=236$~MeV as a function of the c.m. energy. The band corresponds to the SU(2) U$\chi$PT fit. The dashed line shows the $4\pi$ threshold. We do not show two noisy energy levels.}
\label{fig:lattice_fits}
\end{center}
\end{figure}

We determine the two unknown LECs by fitting the 22 energy levels obtained at a single quark mass and spatial volume. In practice, we input the U$\chi$PT amplitude into Eq.~\ref{eq:QC} and compute the spectra for a given set of LECs, $E^{{\rm U}\chi{\rm PT}}(\{\alpha_i\})$. By varying these LECs we minimize the $\chi^2(\{\alpha_i\})$, defined as
\begin{align}
\chi^2(\{\alpha_i\})=\sum_{j,k}
\delta{E}_{j}(\{\alpha_i\})\,
{\mathbb C}^{-1}_{j,k}\,
\delta{E}_{k}(\{\alpha_i\})
\end{align}
where $\delta{E}_{j}(\{\alpha_i\})=\left[E_j^\text{lat}-E_j^{{\rm U}\chi{\rm PT}}(\{\alpha_i\})\right]$, and $\left\{j,k\right\}$ run over all 22 energy levels. As with the energy levels themselves, the elements of the covariance $\mathbb{C}$ matrix were provided by the Hadron Spectrum Collaboration~\cite{Wilson:2015dqa}. The fit results in $\chi^2/ N_\mathrm{d.o.f.} = 1.26$ for SU(2) U$\chi$PT and is shown in Fig.~\ref{fig:lattice_fits} compared to the lattice determined phase shifts. The LECs and correlations are found to be
\begin{equation}
\begin{array}{l}
\alpha_1(770\text{ MeV})=14.7(4)(2)(1)\times10^{-3}\\
\alpha_2(770\text{ MeV})=-28(6)(3)\left(^{01}_{11}\right)\times10^{-3}
\end{array}
\quad
\left[
\begin{array}{lr}
1 & -0.98\\
 & 1
\end{array}
\right]
\label{eq:lecs}
\end{equation}
The first uncertainty is statistical, the second is the systematic due to the determination of the $\pi$ mass and the anisotropy of the lattice~\footnote{The $\pi$ mass was determined in lattice units to be $a_tm_\pi=0.03928(18)$. The anisotropy of that lattice is defined as $\xi=a_s/a_t$ where $a_s$ and $a_t$ are the lattice spacings in the spatial and temporal extents. The anisotropy has been determined to be $\xi=3.4534(61)$.}, and the third is an estimate of the systematic due to the determination of the lattice spacing. The symmetric matrix on the right of the coefficients denotes the statistical correlation between the two. By analytically continuing the scattering amplitude to complex values of $s=(E_{\pi\pi}^{\star})^2$ we obtain a resonance pole on the unphysical sheet, corresponding to taking the negative root when computing the c.m. momentum $q^\star$. At these quark masses, we find a $\rho$ pole at $E_\rho=782(2)-\frac{i}{2}\,85(2)$~MeV with a width, $\Gamma_\rho\equiv -2~{\rm Im}(E_\rho)=85(2)$~MeV. We observe good agreement with the result from the Hadron Spectrum Collaboration where the poles were determined using other parameterizations of the scattering amplitude. This emphasizes the fact that the lattice QCD spectrum properly constrains the scattering phase shift independently of the parameterization chosen. 

\begin{figure}[t]
\begin{center}
\hspace*{-.5cm}                                                           
\includegraphics[scale=0.38]{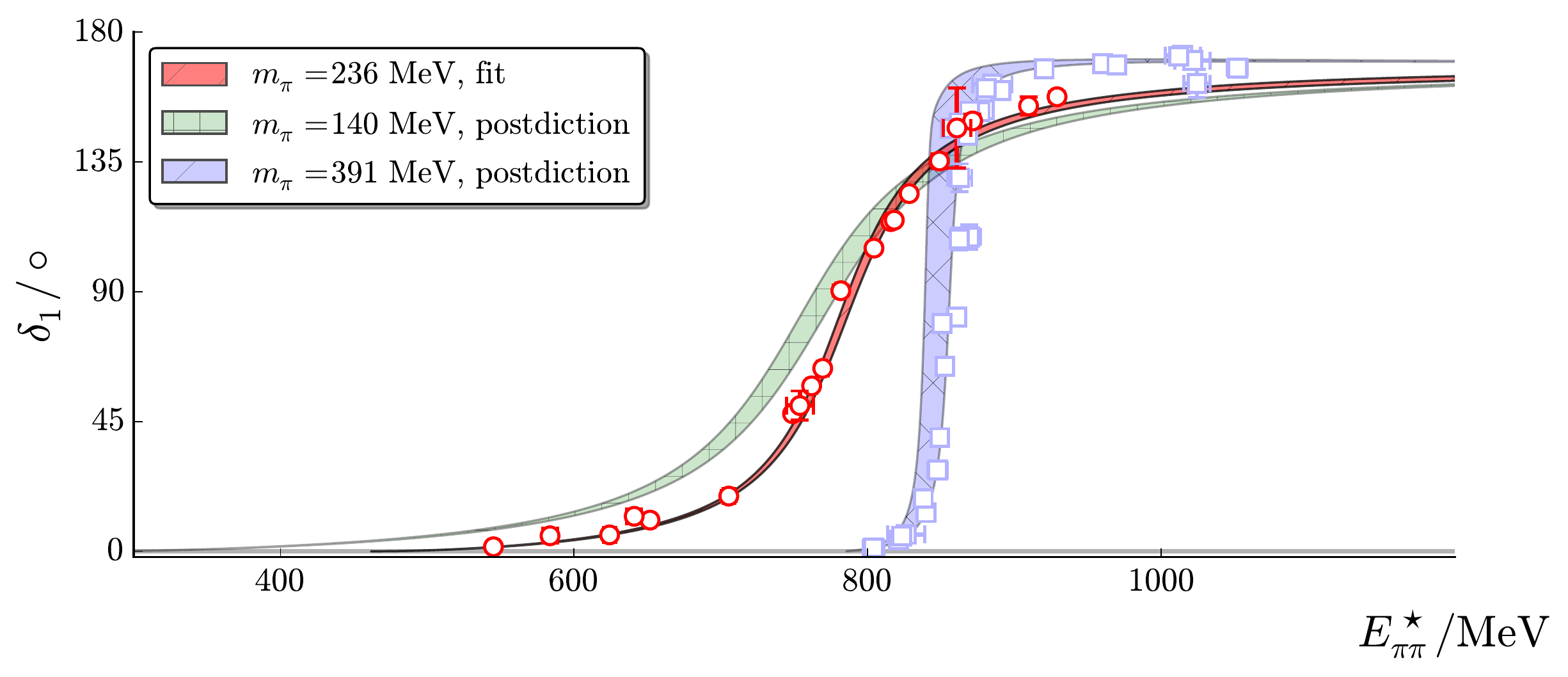}
\caption{$I=1$ $\pi\pi$ phase shifts at three pion masses. In red we show the lattice-determined phase shifts, along with the SU(2) U$\chi$PT fit to the spectrum at $m_\pi= 236$~MeV. The green band shows the extrapolation to the experimental pion mass. In blue we show the discrete points from the lattice calculation at $m_\pi= 391$~MeV~\cite{Dudek:2012xn} and the extrapolation from the parameters determined from this 236 MeV fit. The extrapolated bands include both statistical and systematic errors discussed in the text.  }
\label{fig:extrapolations}
\end{center}
\end{figure}

The power of the U$\chi$PT amplitude is that it allows one to extrapolate these quantities as a function of pion mass. In Fig.~\ref{fig:extrapolations} we show the result of this exercise using the mean values of the coefficients in Eq.~\ref{eq:lecs} and propagating both statistical and systematic uncertainties. We show the \emph{postdiction} for $m_\pi= 140$ MeV and $m_\pi= 391$ MeV, where an earlier calculation also extracted the $\pi\pi$ scattering amplitude containing the $\rho$ resonance~\cite{Dudek:2012xn}. We emphasize that in Ref.~\cite{Pelaez:2010fj} it is clearly explained that U$\chi$PT is not expected to reliably describe lattice QCD results above $m_\pi\sim300-350$~MeV. Despite this formal constraint and the slight deviation at $m_\pi= 391$ MeV from the lattice results, U$\chi$PT produces phase shifts that resemble both experimental and lattice determinations as a function of $m_\pi$.

\begin{figure}[t]
\begin{center}
\hspace*{-.5cm}                                                           
\includegraphics[scale=0.38]{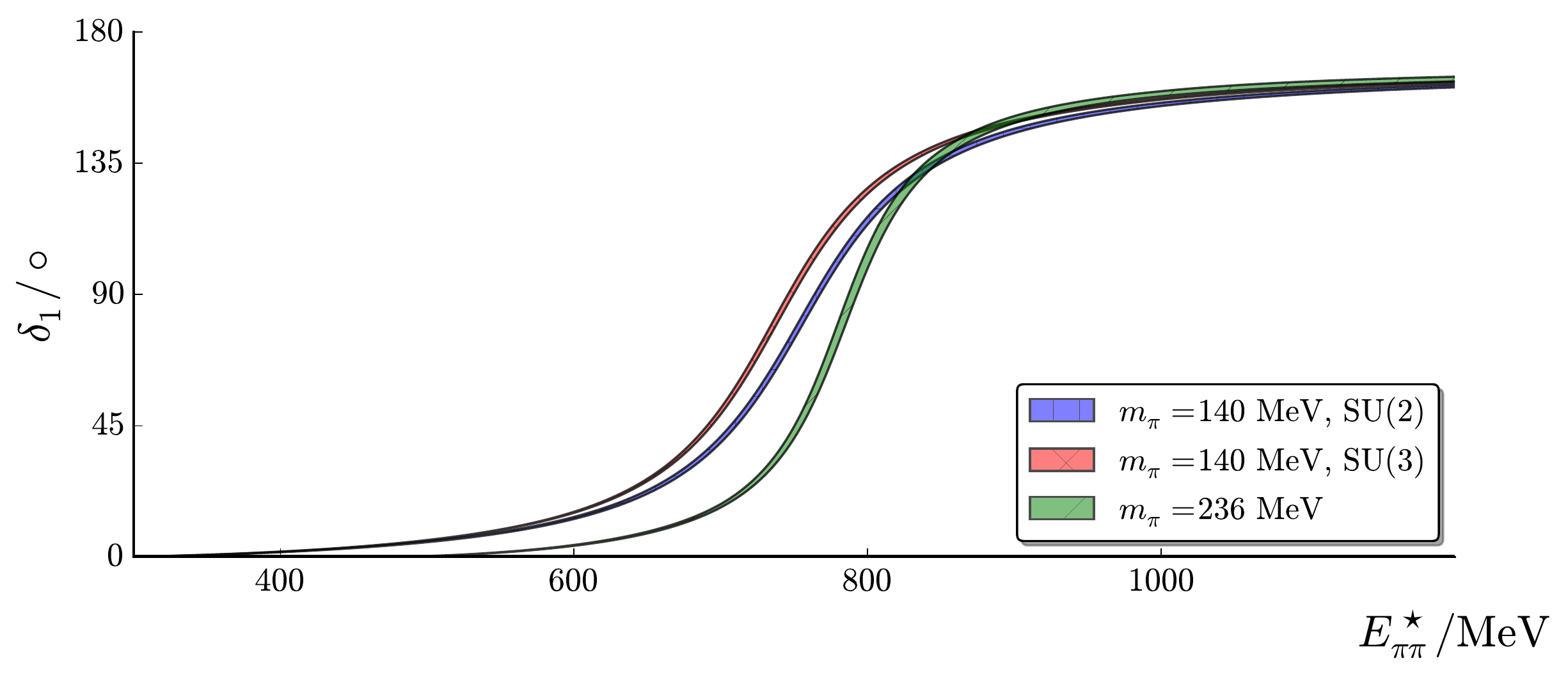}
\caption{ Extrapolations of the phase shift determined at $m_\pi\approx 236$~MeV (green) to the physical point done using SU(2) and SU(3) in blue and red respectively. The extrapolated bands include only statistical error. For an estimate of systematics and a comparison with experimental data, see Fig.~\ref{fig:global_comparison}.}
\label{fig:su2vssu3}
\end{center}
\end{figure}

In Fig.~\ref{fig:su2vssu3} we show a comparison of the results of the extrapolation using SU(2) and SU(3) versions of U$\chi$PT. Given that SU(3)-breaking effects are large, SU(3) $\chi PT$ has a poorer convergence than that the SU(2) counterpart. Therefore, we expect the SU(3) extrapolation to have a significantly larger systematic uncertainty. Assessing such systematic lies outside of the scope of the present work.

\begin{figure*}[t]
\begin{center}
\hspace*{-.5cm}                                                           
\includegraphics[scale=0.53]{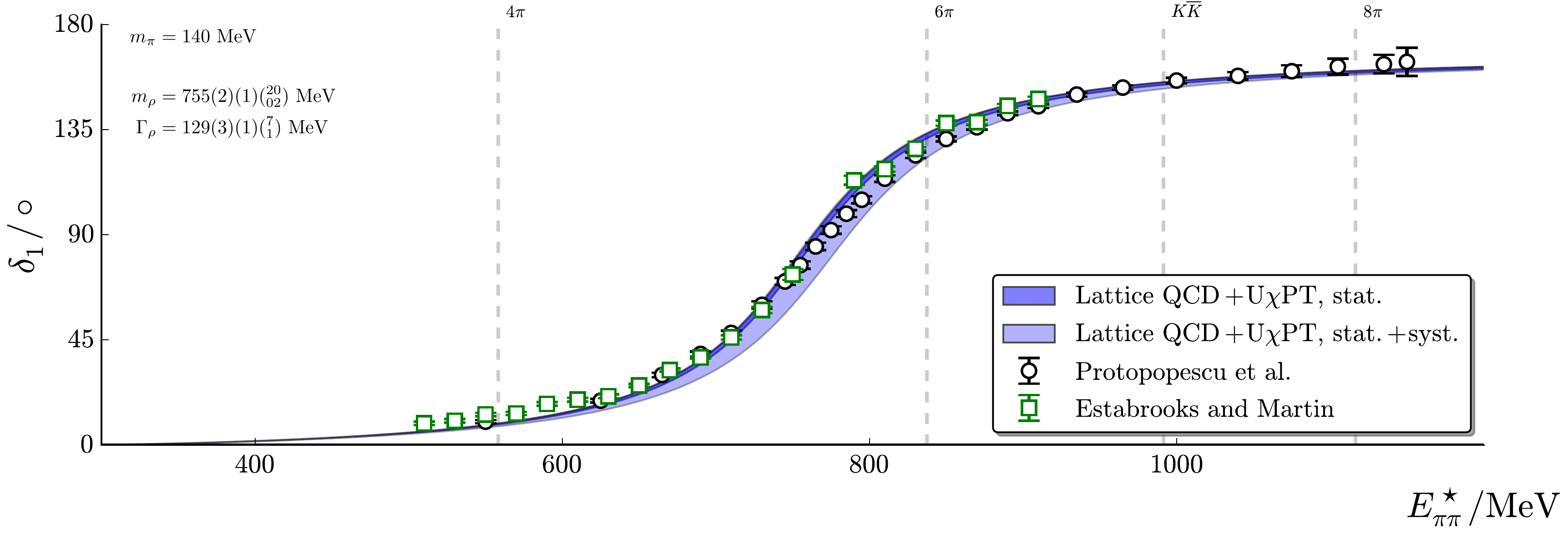}
\caption{Shown is the extrapolation to the physical quark masses of the $(\ell,I)=(1,1)$ $\pi\pi$ scattering phase shift. This is plotted as a function of the c.m. energy $(E^\star_{\pi\pi})$. The darker blue inner band includes only statistical uncertainty, while the lighter outer band also includes systematic uncertainties explained in the text. We see good agreement with the experimental phase shift shown as black circles~\cite{Protopopescu:1973sh} and green squares~\cite{Estabrooks:1974vu}. The dashed lines denote the $4\pi$, $6\pi$, $K\overline{K}$ and $8\pi$ thresholds, which appear to play a negligible role. }\label{fig:global_comparison}
\end{center}
\end{figure*}

In Fig.~\ref{fig:global_comparison} we present our final result for the chiral extrapolation of the $\pi\pi$ phase shift using SU(2) U$\chi$PT. The result includes a propagation of statistical and systematic uncertainties. The largest uncertainty is due to the determination of the lattice spacing, where we aim to be conservative. Overall, we find good agreement with the experimental phase shift~\cite{Protopopescu:1973sh, Estabrooks:1974vu} up to center of mass energies of 1.2~GeV, well above the $4\pi$, $6\pi$, $K\overline{K}$ and $8\pi$ thresholds. By analytically continuing the amplitude into the complex plane, we find a postdiction of the $\rho$ pole at the physical point $E_\rho=\Erho$. 

In order to compare with experimental determinations of the mass and width of the $\rho$, we must restrict out attention to those determinations which have used the model-independent definitions $m_\rho={\rm Re}(E_\rho)$ and $\Gamma_\rho=-2\,{\rm Im}(E_\rho)$. We contrast this with the standard procedure of quoting the mass and width parameters appearing in the Breit-Wigner parametrization of the scattering amplitude (as is done in the Particle Data Group book~\cite{Agashe:2014kda}). Only in the very narrow width limit do these two definitions coincide.

\begin{figure}[t]
\begin{center}
\hspace*{-.5cm}                                                           
\includegraphics[scale=0.28]{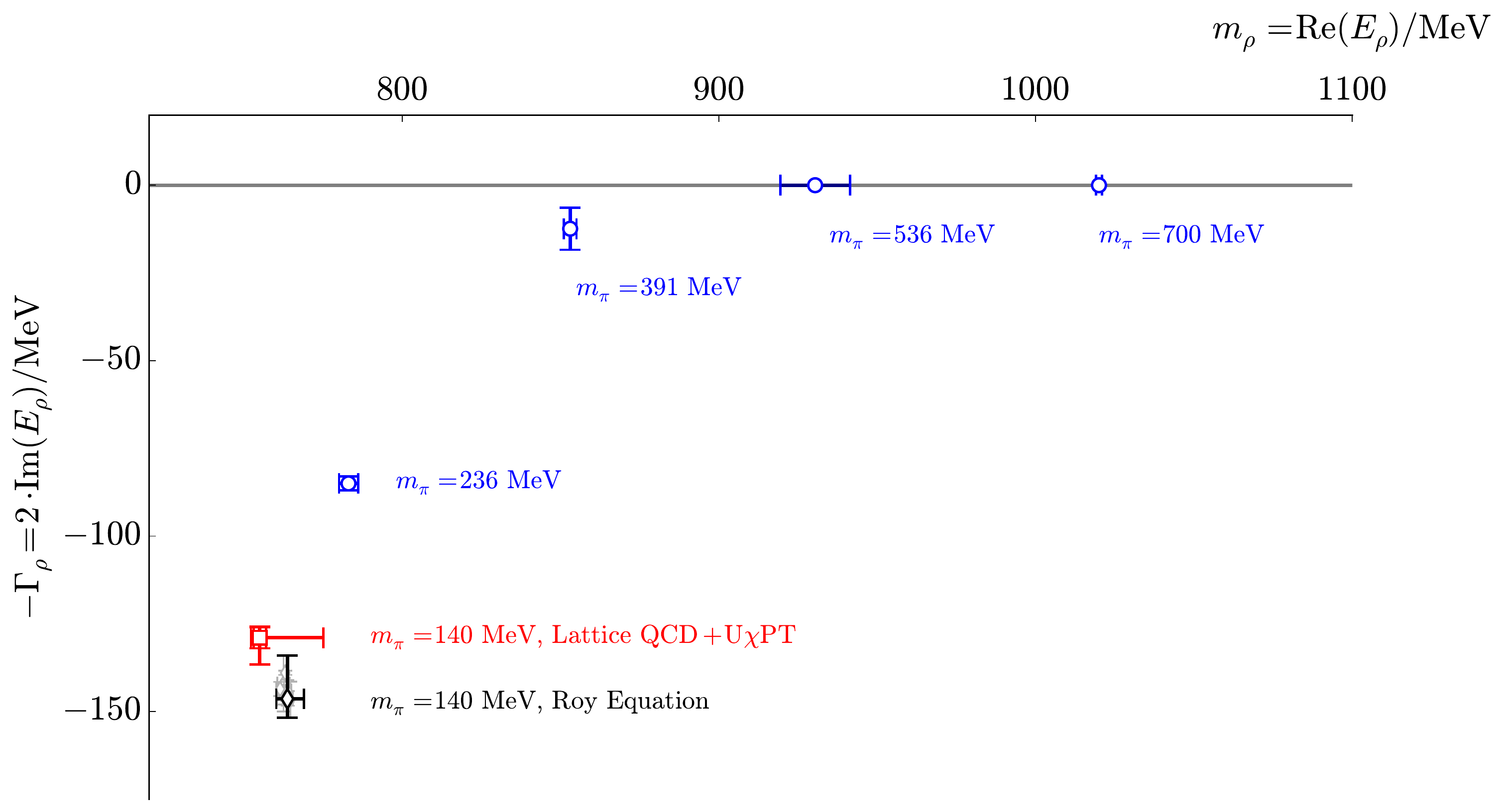}
\caption{We compare our determination of the $\rho$ pole [red square] with previous lattice calculations [blue circles] performed using unphysically heavy quark masses where the $\rho$ is stable~\cite{Lin:2008pr} as well as unstable~\cite{Dudek:2012xn, Wilson:2015dqa}. This is compared with pole determinations obtained from solutions to the Roy equation~\cite{Roy:1971tc} constrained from experimental data [gray diamonds]~\cite{Masjuan:2014psa, Ananthanarayan:2000ht, Colangelo:2001df, Zhou:2004ms, GarciaMartin:2011jx, Masjuan:2013jha}. Using these we highlight a black diamond whose uncertainty is defined to include all determinations from these references up to one standard deviation. 
}\label{fig:rho_mpi}
\end{center}
\end{figure}

In Fig.~\ref{fig:rho_mpi} we show our determination of the $\rho$ pole. For comparison we show those obtained in Refs.~\cite{Masjuan:2014psa, Ananthanarayan:2000ht, Colangelo:2001df, Zhou:2004ms, GarciaMartin:2011jx, Masjuan:2013jha} by solving the Roy equation~\cite{Roy:1971tc} and using experimental data as input. Since these results cover a large area, we highlight a dark point which encompasses all pole positions. Identifying this as an estimate of the overall systematic and statistical uncertainty, we find good agreement with our determination. We also show the pole position obtained in previous lattice QCD calculations~\cite{Lin:2008pr, Dudek:2012xn, Wilson:2015dqa}, including those where the $\rho$ is stable. This plot serves as a nice illustration of the trajectory being taken by the $\rho$ pole as a function of $m_\pi$. For heavy quark masses, the $\rho$ is stable and its pole lies on the real axis. As the quark mass decreases, the $\rho$ becomes unstable and acquires a non-zero width, sending the pole off the real axis.

We compare the LECs determined here with those determined in Refs.~\cite{Hanhart:2008mx, Pelaez:2010fj, Pelaez:2006nj}: $\alpha_1(770~{\rm MeV})\times 10^{3}\in [9,13]$ and $\alpha_2(770~{\rm MeV})\times 10^{3}\in [1,12]$. We observe a qualitative discrepancy between our determination of $\alpha_2$ and those determined in these references. This can be explained by two facts. First, as discussed in Ref.~\cite{Dobado:1996ps}, the $(\ell,I)=(1,1)$ amplitude primarily depends on $\alpha_1$. Second, as mentioned above, the definition and value of these parameters depend on higher order corrections in the chiral expansion~\cite{Pelaez:2010fj}. We suspect that by performing simultaneous fits of various channels while including higher order corrections one will see a convergence of these results. Implementing these techniques for channels including scalar resonances like the $f_0(500)$ would require using the \emph{modified Inverse Amplitude Method} to have the correct analytic structure below threshold~\cite{GomezNicola:2007qj, Pelaez:2010fj}.  The implementation of this awaits the lattice QCD calculation of these channels using $m_\pi=236$~MeV.

\emph{Final remarks:} We present the first extrapolation of a resonant amplitude from lattice QCD. To perform the extrapolation we used U$\chi$PT, an effective field theory that at low-energies coincides with $\chi$PT and at high-energies generates resonances dynamically. In this framework, resonances are manifested naturally as singularities in amplitudes. We observe that this effective field theory does a remarkable job in describing the recent results of the Hadron Spectrum Collaboration. Using the lattice QCD spectrum to constrain the LECs of the theory, we find good agreement with the experimentally measured $\pi\pi$ scattering phase shift up to energies above the $8\pi$ and $K\overline{K}$ thresholds illustrating the significance of this result. We observe the extrapolated amplitude to have a pole, corresponding to the $\rho$ meson, which agrees with previous determinations using dispersive analysis of experimental $\pi\pi$ scattering data.

It is desirable to study more complex systems such as highly energetic exotic hadrons (e.g., the $\pi_1(1400)$ resonance) or heavy meson weak decays (e.g., $D\rightarrow \pi\pi/K\overline{K}$ \cite{Hansen:2012tf, Aaij:2011in}), however it is not yet clear when a finite volume formalism rigorously accommodating all open multiparticle channels will be available. We demonstrate that by properly constraining the scattering amplitude at a value of the pion mass where fewer channels are kinematically open, one can perform an extrapolation to the physical point.
  
These methods may be applied to obtain a wide range of hadron scattering amplitudes that are presently being extracted from lattice QCD, in both the light and heavy quark sectors~\cite{Dudek:2012xn, Dudek:2010ew, Dudek:2012gj, Dudek:2014qha, Wilson:2014cna, Lang:2011mn, Torres:2014vna, Briceno:2015dca}. 
 It is hoped that these concepts could be extended and applied to scattering processes containing highly excited and exotic resonances to gain deeper understanding of QCD and the excited spectrum of hadrons.

\appendix
\section{Chiral Lagrangian and Scattering Amplitude\label{sec:ChPT}}
Here we present the key results of SU(2) $\chi$PT as derived in Ref.~\cite{Gasser:1983kx}. The relevant terms of the leading order (LO) and next-to-leading order (NLO) terms of the chiral Lagrangian (in the isospin limit $m_u=m_d$),
\begin{align}
{\cal L}_{\rm LO}&=\frac{f_0^2}{4}\Tr\left(\partial_\mu U\partial^\mu U^\dagger\right)+\frac{m_0^2\,f_0^2}{4}\Tr\left(U^\dagger+U\right)\nonumber\\
{\cal L}_{\rm NLO}&=\frac{\ell_1}{4}\left[\Tr\left(\partial_\mu U\partial^\mu U^\dagger\right)\right]^2+\frac{\ell_2}{4}\left[\Tr\left(\partial_\mu U\partial_\nu U^\dagger\right)\right]\left[\Tr\left(\partial^\mu U\partial^\nu U^\dagger\right)\right]\nonumber\\
&+\frac{m_0^4\,\ell_3}{16}\left[\Tr\left(U^\dagger+U\right)\right]^2+\frac{m_0^2\,\ell_4}{4}\Tr\left(\partial^2U^\dagger-\partial^2U\right)+\ldots,
\end{align}
are written in terms of the parameters $f_0$ (related to the pion decay constant) and $m_0$ (related to the pion mass), and the matrix of pion fields,
\begin{align}
U&=\exp\left\lbrace\frac{i}{f_0}\begin{pmatrix}\pi^0 & \sqrt{2}\pi^+\\ \sqrt{2}\pi^- & -\pi^0\end{pmatrix}\right\rbrace.
\end{align}
Divergences associated with loops with LO vertices are removed by renormalizing the $\ell_i$ LECs from the NLO Lagrangian and physical quantities depend on the renormalized LECs $\ell_i^r(\mu)$. We use $\mu=770$ MeV in this work. At this order in the chiral expansion, it is convenient to introduce $\mu$-independent expressions for the LECs, $\bar{\ell}_i(m_0)$, that depend on the value of $m_0$,
\begin{equation}
\ell_i^r=\frac{\gamma_i}{32\pi^2}\left[\bar{\ell}_i+\ln\left(m_0^2/\mu^2\right)\right],
\end{equation}
where $\gamma_1=\dfrac{1}{3}$, $\gamma_2=\dfrac{2}{3}$, $\gamma_3=-\dfrac{1}{2}$, and $\gamma_4=2$.

We use the standard NLO expressions~\cite{Gasser:1984gg,Gasser:1983yg,Gasser:1983kx} for the physical pion mass and decay constant,
\begin{align}
m_\pi^2&=m_0^2\left[1-\frac{1}{32\pi^2}\frac{m_0^2}{f_0^2}\bar{\ell}_3(m_0)+\ldots\right],\label{eq:mpi}\\
f_\pi&=f_0\left[1+\frac{1}{16\pi^2}\frac{m_0^2}{f_0^2}\bar{\ell}_4(m_0)+\ldots\right]\label{eq:fpi},
\end{align}
to solve for $m_0$ and $f_0$ perturbatively. To fix $m_0$ we use the value of $m_\pi$ that has been determined on the lattice, $m_\pi^{\rm latt.}$. Since $f_\pi$ has not been determined for these lattices, we resort to fixing $f_\pi$ using the experimental value, $f^{\rm exp.}_\pi$. This approximation forces us to use two different values of $m_\pi$ in our fits. More explicitly, for $m_0$ we use,
\begin{align}
m_0^2
&\approx (m_\pi^{\rm latt.})^2\left[1+\frac{1}{32\pi^2}\frac{(m_\pi^{\rm latt.})^2}{f_0^2}\bar{\ell}_3(m_\pi^{\rm latt.})+\ldots\right]\\
&\approx (m_\pi^{\rm latt.})^2\left[1+\frac{1}{32\pi^2}\frac{(m_\pi^{\rm latt.})^2}{(f^{\rm exp.}_\pi)^2}\bar{\ell}_3(m_\pi^{\rm latt.})+\ldots\right],\label{eq:mpi_approx}
\end{align}
were the ellipses denote corrections that appear at higher orders in the chiral expansion. Similarly, for $f_0$,
\begin{align}
f_0&\approx f_\pi^{\rm exp.}\left[1-\frac{1}{16\pi^2}\frac{(m_\pi^{\rm exp.})^2}{(f_\pi^{\rm exp.})^2}\bar{\ell}_4(m_\pi^{\rm exp.})+\ldots\right].
\end{align}
The amplitudes depend on $1/f_0^2$, which we write here perturbatively
\begin{align}
\frac{1}{f_0^2}&\approx 
\frac{1}{(f_\pi^{\rm exp.})^2}\left[1+\frac{2}{16\pi^2}\frac{(m_\pi^{\rm exp.})^2}{(f_\pi^{\rm exp.})^2}\bar{\ell}_4(m_\pi^{\rm exp.})+\ldots\right].
\end{align}

The scattering amplitude prior to partial-wave projection, ${\cal A}(s,t,u)$, can be written as 
\begin{align}
&{\cal A}_{\rm LO}(s,t,u)=\frac{s-m_\pi^2}{f_\pi^2}\nn\\
&{\cal A}_{\rm NLO}(s,t,u)=\frac{s-m_\pi^2}{f\pi^2}\frac{(m_\pi^{\rm exp.})^2}{8\pi^2f_\pi^2}\bar{\ell}_4(m_\pi^{\rm exp.})-\frac{m_\pi^4}{32\pi^2f_\pi^4}\bar{\ell}_3(m_\pi)\nn\\
&+\frac{1}{6f_\pi^4}\left\lbrace 3(s^2-m_\pi^4)\bar{J}(s)+\left[t(t-u)-2m_\pi^2t+4m_\pi^2u-2m_\pi^4\right]\bar{J}(t)\right.\nn\\
&\hspace{.45in}+\left.\left[u(u-t)-2m_\pi^2u+4m_\pi^2t-2m_\pi^4\right]\bar{J}(u)\right\rbrace\nn\\
&+\frac{1}{96\pi^2f_\pi^4}\left\lbrace 2(\bar{\ell}_1(m_\pi)-4/3)(s-2m_\pi^2)^2\right.\nn\\
&\hspace{.65in}\left.+(\bar{\ell}_2(m_\pi)-5/6)\left[s^2+(t-u)^2\right]-12m_\pi^2s+15m_\pi^4\right\rbrace,
\label{eq:ampChiPT}
\end{align}
where
\begin{equation}
\bar{J}(s)=\frac{1}{16\pi^2}\left\lbrace\sqrt{1-4m_\pi^2/s}\ln\left[\frac{\sqrt{1-4m_\pi^2/s}-1}{\sqrt{1-4m_\pi^2/s}+1}\right]+2\right\rbrace.
\label{eq:Jbar}
\end{equation}
Note that in Eq.~\ref{eq:ampChiPT} we have implemented the perturbative expressions for $m_0^2$ and $\dfrac{1}{f_0^2}$ described above. In Eq.~\ref{eq:ampChiPT} and Eq.~\ref{eq:Jbar} we use the notation $m_\pi=m_\pi^{\rm latt.}$ and $f_\pi=f_\pi^{\rm exp.}$. The amplitude ${\cal A}(s,t,u)$ can then be projected into a partial wave $\ell$ using,
\begin{align}
\mathcal{M}_\ell= \frac{1}{2}\int_{-1}^{1} dz\; P_\ell(z)\; {\cal A}(s,t(s,z),u(s,z))
\end{align}
where $z=\cos\theta$ and $\theta$ is the $s$-channel c.m. frame scattering angle. In this work, we also project onto the $I=1$ channel,
\begin{equation}
{\cal M}^1(s,t,u)={\cal M}(t,s,u)-{\cal M}(u,t,s).
\end{equation}
One can show that the only linear combinations of LECs contributing to the isotriplet scattering amplitude are $\alpha_1\equiv-2\ell^r_1+\ell^r_2$ and $\alpha_2\equiv\ell^r_4$, which are the ones determined in this work.

\section{The Inverse Amplitude Method\label{sec:IAmpMeth}}
Although U$\chi$PT has been extensively discussed in the literature, here we sketch the derivation of Eq.~\ref{eq:MUChPT} presented in Ref.~\cite{GomezNicola:2001as} in an effort to make this article more self-contained. The basic idea, as already mentioned above, is to assure that unitarity is satisfied exactly at each order in the chiral expansion. We begin by giving the standard relation between the $S$-matrix and the partial-wave projected scattering amplitude, $\mathcal{M}$,
\begin{equation}
S=1+2i\sigma \mathcal{M},
\end{equation}
where $\sigma=q/16\pi E_{\pi\pi}^\star$. Unitarity enforces
\begin{equation}
{\rm Im}(\mathcal{M})=\sigma|\mathcal{M}|^2,
\label{eq:OpticalThm}
\end{equation}
which is the familiar Optical Theorem. This condition can be rewritten as
\begin{equation}
{\rm Im}(\mathcal{M}^{-1})=-\sigma, 
\end{equation}
which leads us to
\begin{equation}
\mathcal{M}=({\rm Re}(\mathcal{M}^{-1})-i\sigma)^{-1}.
\label{eq:TwithUnitarity}
\end{equation}

If $\mathcal{M}$ is evaluated perturbatively as detailed in~\ref{sec:ChPT}, $\mathcal{M}=\mathcal{M}_{\rm LO}+\mathcal{M}_{\rm NLO}+\ldots$, we can expand its inverse to find,
\begin{equation}
\mathcal{M}^{-1}=\mathcal{M}^{-1}_{\rm LO}\frac{1}{1+\mathcal{M}_{\rm LO}^{-1}\,\mathcal{M}_{\rm NLO}+\ldots}=\mathcal{M}_{\rm LO}^{-1}\left(1-\mathcal{M}_{\rm LO}^{-1}\,\mathcal{M}_{\rm NLO}+\ldots\right).
\end{equation}
Since $\mathcal{M}_{\rm LO}$ is real,
\begin{equation}
{\rm Re}(\mathcal{M}^{-1})=\mathcal{M}_{\rm LO}^{-1}\left(1-\mathcal{M}_{\rm LO}^{-1}\,{\rm Re}(\mathcal{M}_{\rm NLO})+\ldots\right),
\end{equation}
which we insert into Eq.~\ref{eq:TwithUnitarity} to find,
\begin{align}
\mathcal{M}&=\frac{1}{\mathcal{M}_{\rm LO}^{-1}\left(1-\mathcal{M}_{\rm LO}^{-1}\,{\rm Re}(\mathcal{M}_{\rm NLO})+\ldots\right)-i\sigma}\nonumber\\
&\approx \mathcal{M}_{\rm LO}\frac{1}{\mathcal{M}_{\rm LO}-{\rm Re}(\mathcal{M}_{\rm NLO})-i\sigma \mathcal{M}_{\rm LO}^2}\mathcal{M}_{\rm LO}.
\label{eq:Trewritten}
\end{align}

Finally, let us return to Eq.~\ref{eq:OpticalThm} and enumerate the unitarity constraints order by order,
\begin{align}
{\rm LO}&:\qquad{\rm Im}(\mathcal{M}_{\rm LO})=0\nonumber\\
{\rm NLO}&:\qquad{\rm Im}(\mathcal{M}_{\rm NLO})=\sigma|\mathcal{M}_{\rm LO}|^2=\sigma \mathcal{M}_{\rm LO}^2.
\label{eq:ImT4}
\end{align}
Thus, putting Eq.~\ref{eq:ImT4} into Eq.~\ref{eq:Trewritten}, we reproduce Eq.~\ref{eq:MUChPT}.
 
\subsection*{Acknowledgments}
We thank our colleagues in the Hadron Spectrum Collaboration, in particular J.J. Dudek,  R.G. Edwards and C.E. Thomas, for providing the correlated data sets and for useful discussions and feedback on the manuscript. D.R.B. would like to thank J. Emerick, C. Madrid, and K. Robertson for their help with this project. D.J.W. and R.A.B. would like to thank I. Danilkin and E. Passemar for many useful discussions. R.A.B. acknowledges support from the U.S. Department of Energy contract DE-AC05-06OR23177, under which Jefferson Science Associates, LLC, manages and operates the Jefferson Lab. D.J.W. acknowledges support from the U.S. Department of Energy contract DE-SC0006765.

\bibliographystyle{apsrev}
\bibliography{bibi}

\begin{thebibliography}{76}
\expandafter\ifx\csname natexlab\endcsname\relax\def\natexlab#1{#1}\fi
\expandafter\ifx\csname bibnamefont\endcsname\relax
  \def\bibnamefont#1{#1}\fi
\expandafter\ifx\csname bibfnamefont\endcsname\relax
  \def\bibfnamefont#1{#1}\fi
\expandafter\ifx\csname citenamefont\endcsname\relax
  \def\citenamefont#1{#1}\fi
\expandafter\ifx\csname url\endcsname\relax
  \def\url#1{\texttt{#1}}\fi
\expandafter\ifx\csname urlprefix\endcsname\relax\def\urlprefix{URL }\fi
\providecommand{\bibinfo}[2]{#2}
\providecommand{\eprint}[2][]{\url{#2}}

\bibitem[{\citenamefont{Borsanyi et~al.}(2015)}]{Borsanyi:2014jba}
\bibinfo{author}{\bibfnamefont{S.}~\bibnamefont{Borsanyi}}
  \bibnamefont{et~al.}, \bibinfo{journal}{Science}
  \textbf{\bibinfo{volume}{347}}, \bibinfo{pages}{1452} (\bibinfo{year}{2015}),
  \eprint{1406.4088}.

\bibitem[{\citenamefont{Borsanyi et~al.}(2013)}]{Borsanyi:2013lga}
\bibinfo{author}{\bibfnamefont{S.}~\bibnamefont{Borsanyi}} \bibnamefont{et~al.}
  (\bibinfo{collaboration}{Budapest-Marseille-Wuppertal}),
  \bibinfo{journal}{Phys.Rev.Lett.} \textbf{\bibinfo{volume}{111}},
  \bibinfo{pages}{252001} (\bibinfo{year}{2013}), \eprint{1306.2287}.

\bibitem[{\citenamefont{Aoki et~al.}(2012)}]{Aoki:2012st}
\bibinfo{author}{\bibfnamefont{S.}~\bibnamefont{Aoki}} \bibnamefont{et~al.},
  \bibinfo{journal}{Phys. Rev.} \textbf{\bibinfo{volume}{D86}},
  \bibinfo{pages}{034507} (\bibinfo{year}{2012}), \eprint{1205.2961}.

\bibitem[{\citenamefont{Briceno
  et~al.}(2015{\natexlab{a}})\citenamefont{Briceno, Davoudi, and
  Luu}}]{Briceno:2014tqa}
\bibinfo{author}{\bibfnamefont{R.~A.} \bibnamefont{Briceno}},
  \bibinfo{author}{\bibfnamefont{Z.}~\bibnamefont{Davoudi}}, \bibnamefont{and}
  \bibinfo{author}{\bibfnamefont{T.~C.} \bibnamefont{Luu}},
  \bibinfo{journal}{J. Phys.} \textbf{\bibinfo{volume}{G42}},
  \bibinfo{pages}{023101} (\bibinfo{year}{2015}{\natexlab{a}}),
  \eprint{1406.5673}.

\bibitem[{\citenamefont{Briceno}(2015)}]{Briceno:2014pka}
\bibinfo{author}{\bibfnamefont{R.~A.} \bibnamefont{Briceno}},
  \bibinfo{journal}{PoS} \textbf{\bibinfo{volume}{LATTICE2014}},
  \bibinfo{pages}{008} (\bibinfo{year}{2015}), \eprint{1411.6944}.

\bibitem[{\citenamefont{Yamazaki}(2015)}]{Yamazaki:2015nka}
\bibinfo{author}{\bibfnamefont{T.}~\bibnamefont{Yamazaki}},
  \bibinfo{journal}{PoS} \textbf{\bibinfo{volume}{LATTICE2014}},
  \bibinfo{pages}{009} (\bibinfo{year}{2015}), \eprint{1503.08671}.

\bibitem[{\citenamefont{Prelovsek}(2014)}]{Prelovsek:2014zga}
\bibinfo{author}{\bibfnamefont{S.}~\bibnamefont{Prelovsek}},
  \bibinfo{journal}{PoS} \textbf{\bibinfo{volume}{LATTICE2014}},
  \bibinfo{pages}{015} (\bibinfo{year}{2014}), \eprint{1411.0405}.

\bibitem[{\citenamefont{Wilson et~al.}(2015{\natexlab{a}})\citenamefont{Wilson,
  Briceno, Dudek, Edwards, and Thomas}}]{Wilson:2015dqa}
\bibinfo{author}{\bibfnamefont{D.~J.} \bibnamefont{Wilson}},
  \bibinfo{author}{\bibfnamefont{R.~A.} \bibnamefont{Briceno}},
  \bibinfo{author}{\bibfnamefont{J.~J.} \bibnamefont{Dudek}},
  \bibinfo{author}{\bibfnamefont{R.~G.} \bibnamefont{Edwards}},
  \bibnamefont{and} \bibinfo{author}{\bibfnamefont{C.~E.}
  \bibnamefont{Thomas}}, \bibinfo{journal}{Phys. Rev.}
  \textbf{\bibinfo{volume}{D92}}, \bibinfo{pages}{094502}
  (\bibinfo{year}{2015}{\natexlab{a}}), \eprint{1507.02599}.

\bibitem[{\citenamefont{Luscher}(1986)}]{Luscher:1986pf}
\bibinfo{author}{\bibfnamefont{M.}~\bibnamefont{Luscher}},
  \bibinfo{journal}{Commun. Math. Phys.} \textbf{\bibinfo{volume}{105}},
  \bibinfo{pages}{153} (\bibinfo{year}{1986}).

\bibitem[{\citenamefont{Luscher}(1991)}]{Luscher:1990ux}
\bibinfo{author}{\bibfnamefont{M.}~\bibnamefont{Luscher}},
  \bibinfo{journal}{Nucl. Phys.} \textbf{\bibinfo{volume}{B354}},
  \bibinfo{pages}{531} (\bibinfo{year}{1991}).

\bibitem[{\citenamefont{He et~al.}(2005)\citenamefont{He, Feng, and
  Liu}}]{He:2005ey}
\bibinfo{author}{\bibfnamefont{S.}~\bibnamefont{He}},
  \bibinfo{author}{\bibfnamefont{X.}~\bibnamefont{Feng}}, \bibnamefont{and}
  \bibinfo{author}{\bibfnamefont{C.}~\bibnamefont{Liu}},
  \bibinfo{journal}{JHEP} \textbf{\bibinfo{volume}{07}}, \bibinfo{pages}{011}
  (\bibinfo{year}{2005}), \eprint{hep-lat/0504019}.

\bibitem[{\citenamefont{Briceno and
  Davoudi}(2013{\natexlab{a}})}]{Briceno:2012yi}
\bibinfo{author}{\bibfnamefont{R.~A.} \bibnamefont{Briceno}} \bibnamefont{and}
  \bibinfo{author}{\bibfnamefont{Z.}~\bibnamefont{Davoudi}},
  \bibinfo{journal}{Phys. Rev.} \textbf{\bibinfo{volume}{D88}},
  \bibinfo{pages}{094507} (\bibinfo{year}{2013}{\natexlab{a}}),
  \eprint{1204.1110}.

\bibitem[{\citenamefont{Hansen and Sharpe}(2012)}]{Hansen:2012tf}
\bibinfo{author}{\bibfnamefont{M.~T.} \bibnamefont{Hansen}} \bibnamefont{and}
  \bibinfo{author}{\bibfnamefont{S.~R.} \bibnamefont{Sharpe}},
  \bibinfo{journal}{Phys. Rev.} \textbf{\bibinfo{volume}{D86}},
  \bibinfo{pages}{016007} (\bibinfo{year}{2012}), \eprint{1204.0826}.

\bibitem[{\citenamefont{Briceno}(2014)}]{Briceno:2014oea}
\bibinfo{author}{\bibfnamefont{R.~A.} \bibnamefont{Briceno}},
  \bibinfo{journal}{Phys. Rev.} \textbf{\bibinfo{volume}{D89}},
  \bibinfo{pages}{074507} (\bibinfo{year}{2014}), \eprint{1401.3312}.

\bibitem[{\citenamefont{Dudek et~al.}(2014)\citenamefont{Dudek, Edwards,
  Thomas, and Wilson}}]{Dudek:2014qha}
\bibinfo{author}{\bibfnamefont{J.~J.} \bibnamefont{Dudek}},
  \bibinfo{author}{\bibfnamefont{R.~G.} \bibnamefont{Edwards}},
  \bibinfo{author}{\bibfnamefont{C.~E.} \bibnamefont{Thomas}},
  \bibnamefont{and} \bibinfo{author}{\bibfnamefont{D.~J.} \bibnamefont{Wilson}}
  (\bibinfo{collaboration}{Hadron Spectrum}), \bibinfo{journal}{Phys. Rev.
  Lett.} \textbf{\bibinfo{volume}{113}}, \bibinfo{pages}{182001}
  (\bibinfo{year}{2014}), \eprint{1406.4158}.

\bibitem[{\citenamefont{Wilson et~al.}(2015{\natexlab{b}})\citenamefont{Wilson,
  Dudek, Edwards, and Thomas}}]{Wilson:2014cna}
\bibinfo{author}{\bibfnamefont{D.~J.} \bibnamefont{Wilson}},
  \bibinfo{author}{\bibfnamefont{J.~J.} \bibnamefont{Dudek}},
  \bibinfo{author}{\bibfnamefont{R.~G.} \bibnamefont{Edwards}},
  \bibnamefont{and} \bibinfo{author}{\bibfnamefont{C.~E.}
  \bibnamefont{Thomas}}, \bibinfo{journal}{Phys. Rev.}
  \textbf{\bibinfo{volume}{D91}}, \bibinfo{pages}{054008}
  (\bibinfo{year}{2015}{\natexlab{b}}), \eprint{1411.2004}.

\bibitem[{\citenamefont{Hansen and Sharpe}(2014)}]{Hansen:2014eka}
\bibinfo{author}{\bibfnamefont{M.~T.} \bibnamefont{Hansen}} \bibnamefont{and}
  \bibinfo{author}{\bibfnamefont{S.~R.} \bibnamefont{Sharpe}},
  \bibinfo{journal}{Phys. Rev.} \textbf{\bibinfo{volume}{D90}},
  \bibinfo{pages}{116003} (\bibinfo{year}{2014}), \eprint{1408.5933}.

\bibitem[{\citenamefont{Hansen and Sharpe}(2015)}]{Hansen:2015zga}
\bibinfo{author}{\bibfnamefont{M.~T.} \bibnamefont{Hansen}} \bibnamefont{and}
  \bibinfo{author}{\bibfnamefont{S.~R.} \bibnamefont{Sharpe}},
  \bibinfo{journal}{Phys. Rev.} \textbf{\bibinfo{volume}{D92}},
  \bibinfo{pages}{114509} (\bibinfo{year}{2015}), \eprint{1504.04248}.

\bibitem[{\citenamefont{Briceno and
  Davoudi}(2013{\natexlab{b}})}]{Briceno:2012rv}
\bibinfo{author}{\bibfnamefont{R.~A.} \bibnamefont{Briceno}} \bibnamefont{and}
  \bibinfo{author}{\bibfnamefont{Z.}~\bibnamefont{Davoudi}},
  \bibinfo{journal}{Phys.Rev.} \textbf{\bibinfo{volume}{D87}},
  \bibinfo{pages}{094507} (\bibinfo{year}{2013}{\natexlab{b}}),
  \eprint{1212.3398}.

\bibitem[{\citenamefont{Polejaeva and Rusetsky}(2012)}]{Polejaeva:2012ut}
\bibinfo{author}{\bibfnamefont{K.}~\bibnamefont{Polejaeva}} \bibnamefont{and}
  \bibinfo{author}{\bibfnamefont{A.}~\bibnamefont{Rusetsky}},
  \bibinfo{journal}{Eur. Phys. J.} \textbf{\bibinfo{volume}{A48}},
  \bibinfo{pages}{67} (\bibinfo{year}{2012}), \eprint{1203.1241}.

\bibitem[{\citenamefont{Dudek et~al.}(2009)\citenamefont{Dudek, Edwards,
  Peardon, Richards, and Thomas}}]{Dudek:2009qf}
\bibinfo{author}{\bibfnamefont{J.~J.} \bibnamefont{Dudek}},
  \bibinfo{author}{\bibfnamefont{R.~G.} \bibnamefont{Edwards}},
  \bibinfo{author}{\bibfnamefont{M.~J.} \bibnamefont{Peardon}},
  \bibinfo{author}{\bibfnamefont{D.~G.} \bibnamefont{Richards}},
  \bibnamefont{and} \bibinfo{author}{\bibfnamefont{C.~E.}
  \bibnamefont{Thomas}}, \bibinfo{journal}{Phys. Rev. Lett.}
  \textbf{\bibinfo{volume}{103}}, \bibinfo{pages}{262001}
  (\bibinfo{year}{2009}), \eprint{0909.0200}.

\bibitem[{\citenamefont{Dudek et~al.}(2011{\natexlab{a}})\citenamefont{Dudek,
  Edwards, Joo, Peardon, Richards, and Thomas}}]{Dudek:2011tt}
\bibinfo{author}{\bibfnamefont{J.~J.} \bibnamefont{Dudek}},
  \bibinfo{author}{\bibfnamefont{R.~G.} \bibnamefont{Edwards}},
  \bibinfo{author}{\bibfnamefont{B.}~\bibnamefont{Joo}},
  \bibinfo{author}{\bibfnamefont{M.~J.} \bibnamefont{Peardon}},
  \bibinfo{author}{\bibfnamefont{D.~G.} \bibnamefont{Richards}},
  \bibnamefont{and} \bibinfo{author}{\bibfnamefont{C.~E.}
  \bibnamefont{Thomas}}, \bibinfo{journal}{Phys. Rev.}
  \textbf{\bibinfo{volume}{D83}}, \bibinfo{pages}{111502}
  (\bibinfo{year}{2011}{\natexlab{a}}), \eprint{1102.4299}.

\bibitem[{\citenamefont{Liu et~al.}(2012)\citenamefont{Liu, Moir, Peardon,
  Ryan, Thomas, Vilaseca, Dudek, Edwards, Joo, and Richards}}]{Liu:2012ze}
\bibinfo{author}{\bibfnamefont{L.}~\bibnamefont{Liu}},
  \bibinfo{author}{\bibfnamefont{G.}~\bibnamefont{Moir}},
  \bibinfo{author}{\bibfnamefont{M.}~\bibnamefont{Peardon}},
  \bibinfo{author}{\bibfnamefont{S.~M.} \bibnamefont{Ryan}},
  \bibinfo{author}{\bibfnamefont{C.~E.} \bibnamefont{Thomas}},
  \bibinfo{author}{\bibfnamefont{P.}~\bibnamefont{Vilaseca}},
  \bibinfo{author}{\bibfnamefont{J.~J.} \bibnamefont{Dudek}},
  \bibinfo{author}{\bibfnamefont{R.~G.} \bibnamefont{Edwards}},
  \bibinfo{author}{\bibfnamefont{B.}~\bibnamefont{Joo}}, \bibnamefont{and}
  \bibinfo{author}{\bibfnamefont{D.~G.} \bibnamefont{Richards}}
  (\bibinfo{collaboration}{Hadron Spectrum}), \bibinfo{journal}{JHEP}
  \textbf{\bibinfo{volume}{07}}, \bibinfo{pages}{126} (\bibinfo{year}{2012}),
  \eprint{1204.5425}.

\bibitem[{\citenamefont{Aaij et~al.}(2012)}]{Aaij:2011in}
\bibinfo{author}{\bibfnamefont{R.}~\bibnamefont{Aaij}} \bibnamefont{et~al.}
  (\bibinfo{collaboration}{LHCb}), \bibinfo{journal}{Phys. Rev. Lett.}
  \textbf{\bibinfo{volume}{108}}, \bibinfo{pages}{111602}
  (\bibinfo{year}{2012}), \eprint{1112.0938}.

\bibitem[{\citenamefont{Metivet}(2015)}]{Metivet:2014bga}
\bibinfo{author}{\bibfnamefont{T.}~\bibnamefont{Metivet}}
  (\bibinfo{collaboration}{Budapest-Marseille-Wuppertal}),
  \bibinfo{journal}{PoS} \textbf{\bibinfo{volume}{LATTICE2014}},
  \bibinfo{pages}{079} (\bibinfo{year}{2015}), \eprint{1410.8447}.

\bibitem[{\citenamefont{Dudek et~al.}(2013)\citenamefont{Dudek, Edwards, and
  Thomas}}]{Dudek:2012xn}
\bibinfo{author}{\bibfnamefont{J.~J.} \bibnamefont{Dudek}},
  \bibinfo{author}{\bibfnamefont{R.~G.} \bibnamefont{Edwards}},
  \bibnamefont{and} \bibinfo{author}{\bibfnamefont{C.~E.} \bibnamefont{Thomas}}
  (\bibinfo{collaboration}{Hadron Spectrum}), \bibinfo{journal}{Phys.Rev.}
  \textbf{\bibinfo{volume}{D87}}, \bibinfo{pages}{034505}
  (\bibinfo{year}{2013}), \eprint{1212.0830}.

\bibitem[{\citenamefont{Feng et~al.}(2011)\citenamefont{Feng, Jansen, and
  Renner}}]{Feng:2010es}
\bibinfo{author}{\bibfnamefont{X.}~\bibnamefont{Feng}},
  \bibinfo{author}{\bibfnamefont{K.}~\bibnamefont{Jansen}}, \bibnamefont{and}
  \bibinfo{author}{\bibfnamefont{D.~B.} \bibnamefont{Renner}},
  \bibinfo{journal}{Phys. Rev.} \textbf{\bibinfo{volume}{D83}},
  \bibinfo{pages}{094505} (\bibinfo{year}{2011}), \eprint{1011.5288}.

\bibitem[{\citenamefont{Lang et~al.}(2011)\citenamefont{Lang, Mohler,
  Prelovsek, and Vidmar}}]{Lang:2011mn}
\bibinfo{author}{\bibfnamefont{C.~B.} \bibnamefont{Lang}},
  \bibinfo{author}{\bibfnamefont{D.}~\bibnamefont{Mohler}},
  \bibinfo{author}{\bibfnamefont{S.}~\bibnamefont{Prelovsek}},
  \bibnamefont{and} \bibinfo{author}{\bibfnamefont{M.}~\bibnamefont{Vidmar}},
  \bibinfo{journal}{Phys. Rev.} \textbf{\bibinfo{volume}{D84}},
  \bibinfo{pages}{054503} (\bibinfo{year}{2011}), \bibinfo{note}{[Erratum:
  Phys. Rev.D89,no.5,059903(2014)]}, \eprint{1105.5636}.

\bibitem[{\citenamefont{Pelissier and Alexandru}(2013)}]{Pelissier:2012pi}
\bibinfo{author}{\bibfnamefont{C.}~\bibnamefont{Pelissier}} \bibnamefont{and}
  \bibinfo{author}{\bibfnamefont{A.}~\bibnamefont{Alexandru}},
  \bibinfo{journal}{Phys. Rev.} \textbf{\bibinfo{volume}{D87}},
  \bibinfo{pages}{014503} (\bibinfo{year}{2013}), \eprint{1211.0092}.

\bibitem[{\citenamefont{Aoki et~al.}(2011)}]{Aoki:2011yj}
\bibinfo{author}{\bibfnamefont{S.}~\bibnamefont{Aoki}} \bibnamefont{et~al.}
  (\bibinfo{collaboration}{CS}), \bibinfo{journal}{Phys. Rev.}
  \textbf{\bibinfo{volume}{D84}}, \bibinfo{pages}{094505}
  (\bibinfo{year}{2011}), \eprint{1106.5365}.

\bibitem[{\citenamefont{Aoki et~al.}(2007)}]{Aoki:2007rd}
\bibinfo{author}{\bibfnamefont{S.}~\bibnamefont{Aoki}} \bibnamefont{et~al.}
  (\bibinfo{collaboration}{CP-PACS}), \bibinfo{journal}{Phys. Rev.}
  \textbf{\bibinfo{volume}{D76}}, \bibinfo{pages}{094506}
  (\bibinfo{year}{2007}), \eprint{0708.3705}.

\bibitem[{\citenamefont{Olive et~al.}(2014)}]{Agashe:2014kda}
\bibinfo{author}{\bibfnamefont{K.~A.} \bibnamefont{Olive}} \bibnamefont{et~al.}
  (\bibinfo{collaboration}{Particle Data Group}), \bibinfo{journal}{Chin.
  Phys.} \textbf{\bibinfo{volume}{C38}}, \bibinfo{pages}{090001}
  (\bibinfo{year}{2014}).

\bibitem[{\citenamefont{Oller et~al.}(1998)\citenamefont{Oller, Oset, and
  Pelaez}}]{Oller:1997ng}
\bibinfo{author}{\bibfnamefont{J.~A.} \bibnamefont{Oller}},
  \bibinfo{author}{\bibfnamefont{E.}~\bibnamefont{Oset}}, \bibnamefont{and}
  \bibinfo{author}{\bibfnamefont{J.~R.} \bibnamefont{Pelaez}},
  \bibinfo{journal}{Phys. Rev. Lett.} \textbf{\bibinfo{volume}{80}},
  \bibinfo{pages}{3452} (\bibinfo{year}{1998}), \eprint{hep-ph/9803242}.

\bibitem[{\citenamefont{Dobado and Pelaez}(1997)}]{Dobado:1996ps}
\bibinfo{author}{\bibfnamefont{A.}~\bibnamefont{Dobado}} \bibnamefont{and}
  \bibinfo{author}{\bibfnamefont{J.~R.} \bibnamefont{Pelaez}},
  \bibinfo{journal}{Phys. Rev.} \textbf{\bibinfo{volume}{D56}},
  \bibinfo{pages}{3057} (\bibinfo{year}{1997}), \eprint{hep-ph/9604416}.

\bibitem[{\citenamefont{Oller et~al.}(1999)\citenamefont{Oller, Oset, and
  Pelaez}}]{Oller:1998hw}
\bibinfo{author}{\bibfnamefont{J.~A.} \bibnamefont{Oller}},
  \bibinfo{author}{\bibfnamefont{E.}~\bibnamefont{Oset}}, \bibnamefont{and}
  \bibinfo{author}{\bibfnamefont{J.~R.} \bibnamefont{Pelaez}},
  \bibinfo{journal}{Phys. Rev.} \textbf{\bibinfo{volume}{D59}},
  \bibinfo{pages}{074001} (\bibinfo{year}{1999}), \bibinfo{note}{[Erratum:
  Phys. Rev.D75,099903(2007)]}, \eprint{hep-ph/9804209}.

\bibitem[{\citenamefont{Gomez~Nicola and Pelaez}(2002)}]{GomezNicola:2001as}
\bibinfo{author}{\bibfnamefont{A.}~\bibnamefont{Gomez~Nicola}}
  \bibnamefont{and} \bibinfo{author}{\bibfnamefont{J.~R.}
  \bibnamefont{Pelaez}}, \bibinfo{journal}{Phys. Rev.}
  \textbf{\bibinfo{volume}{D65}}, \bibinfo{pages}{054009}
  (\bibinfo{year}{2002}), \eprint{hep-ph/0109056}.

\bibitem[{\citenamefont{Pelaez and Rios}(2006)}]{Pelaez:2006nj}
\bibinfo{author}{\bibfnamefont{J.~R.} \bibnamefont{Pelaez}} \bibnamefont{and}
  \bibinfo{author}{\bibfnamefont{G.}~\bibnamefont{Rios}},
  \bibinfo{journal}{Phys. Rev. Lett.} \textbf{\bibinfo{volume}{97}},
  \bibinfo{pages}{242002} (\bibinfo{year}{2006}), \eprint{hep-ph/0610397}.

\bibitem[{\citenamefont{Chen and Oset}(2013)}]{Chen:2012rp}
\bibinfo{author}{\bibfnamefont{H.-X.} \bibnamefont{Chen}} \bibnamefont{and}
  \bibinfo{author}{\bibfnamefont{E.}~\bibnamefont{Oset}},
  \bibinfo{journal}{Phys. Rev.} \textbf{\bibinfo{volume}{D87}},
  \bibinfo{pages}{016014} (\bibinfo{year}{2013}), \eprint{1202.2787}.

\bibitem[{\citenamefont{Doring et~al.}(2012)\citenamefont{Doring, Meissner,
  Oset, and Rusetsky}}]{Doring:2012eu}
\bibinfo{author}{\bibfnamefont{M.}~\bibnamefont{Doring}},
  \bibinfo{author}{\bibfnamefont{U.~G.} \bibnamefont{Meissner}},
  \bibinfo{author}{\bibfnamefont{E.}~\bibnamefont{Oset}}, \bibnamefont{and}
  \bibinfo{author}{\bibfnamefont{A.}~\bibnamefont{Rusetsky}},
  \bibinfo{journal}{Eur. Phys. J.} \textbf{\bibinfo{volume}{A48}},
  \bibinfo{pages}{114} (\bibinfo{year}{2012}), \eprint{1205.4838}.

\bibitem[{\citenamefont{Doring and Meissner}(2012)}]{Doring:2011nd}
\bibinfo{author}{\bibfnamefont{M.}~\bibnamefont{Doring}} \bibnamefont{and}
  \bibinfo{author}{\bibfnamefont{U.~G.} \bibnamefont{Meissner}},
  \bibinfo{journal}{JHEP} \textbf{\bibinfo{volume}{01}}, \bibinfo{pages}{009}
  (\bibinfo{year}{2012}), \eprint{1111.0616}.

\bibitem[{\citenamefont{Doring et~al.}(2011)\citenamefont{Doring, Meissner,
  Oset, and Rusetsky}}]{Doring:2011vk}
\bibinfo{author}{\bibfnamefont{M.}~\bibnamefont{Doring}},
  \bibinfo{author}{\bibfnamefont{U.-G.} \bibnamefont{Meissner}},
  \bibinfo{author}{\bibfnamefont{E.}~\bibnamefont{Oset}}, \bibnamefont{and}
  \bibinfo{author}{\bibfnamefont{A.}~\bibnamefont{Rusetsky}},
  \bibinfo{journal}{Eur. Phys. J.} \textbf{\bibinfo{volume}{A47}},
  \bibinfo{pages}{139} (\bibinfo{year}{2011}), \eprint{1107.3988}.

\bibitem[{\citenamefont{Bernard et~al.}(2011)\citenamefont{Bernard, Lage,
  Meissner, and Rusetsky}}]{Bernard:2010fp}
\bibinfo{author}{\bibfnamefont{V.}~\bibnamefont{Bernard}},
  \bibinfo{author}{\bibfnamefont{M.}~\bibnamefont{Lage}},
  \bibinfo{author}{\bibfnamefont{U.~G.} \bibnamefont{Meissner}},
  \bibnamefont{and} \bibinfo{author}{\bibfnamefont{A.}~\bibnamefont{Rusetsky}},
  \bibinfo{journal}{JHEP} \textbf{\bibinfo{volume}{01}}, \bibinfo{pages}{019}
  (\bibinfo{year}{2011}), \eprint{1010.6018}.

\bibitem[{\citenamefont{Nebreda et~al.}(2011)\citenamefont{Nebreda, Pelaez, and
  Rios}}]{Nebreda:2011di}
\bibinfo{author}{\bibfnamefont{J.}~\bibnamefont{Nebreda}},
  \bibinfo{author}{\bibfnamefont{J.~R.} \bibnamefont{Pelaez}},
  \bibnamefont{and} \bibinfo{author}{\bibfnamefont{G.}~\bibnamefont{Rios}},
  \bibinfo{journal}{Phys. Rev.} \textbf{\bibinfo{volume}{D83}},
  \bibinfo{pages}{094011} (\bibinfo{year}{2011}), \eprint{1101.2171}.

\bibitem[{\citenamefont{Rios et~al.}(2008)\citenamefont{Rios, Gomez~Nicola,
  Hanhart, and Pelaez}}]{Rios:2008zr}
\bibinfo{author}{\bibfnamefont{G.}~\bibnamefont{Rios}},
  \bibinfo{author}{\bibfnamefont{A.}~\bibnamefont{Gomez~Nicola}},
  \bibinfo{author}{\bibfnamefont{C.}~\bibnamefont{Hanhart}}, \bibnamefont{and}
  \bibinfo{author}{\bibfnamefont{J.~R.} \bibnamefont{Pelaez}},
  \bibinfo{journal}{AIP Conf. Proc.} \textbf{\bibinfo{volume}{1030}},
  \bibinfo{pages}{268} (\bibinfo{year}{2008}), \eprint{0803.4318}.

\bibitem[{\citenamefont{Guo et~al.}(2009)\citenamefont{Guo, Hanhart,
  Llanes-Estrada, and Meissner}}]{Guo:2008nc}
\bibinfo{author}{\bibfnamefont{F.-K.} \bibnamefont{Guo}},
  \bibinfo{author}{\bibfnamefont{C.}~\bibnamefont{Hanhart}},
  \bibinfo{author}{\bibfnamefont{F.~J.} \bibnamefont{Llanes-Estrada}},
  \bibnamefont{and} \bibinfo{author}{\bibfnamefont{U.-G.}
  \bibnamefont{Meissner}}, \bibinfo{journal}{Phys. Lett.}
  \textbf{\bibinfo{volume}{B678}}, \bibinfo{pages}{90} (\bibinfo{year}{2009}),
  \eprint{0812.3270}.

\bibitem[{\citenamefont{Hanhart et~al.}(2008)\citenamefont{Hanhart, Pelaez, and
  Rios}}]{Hanhart:2008mx}
\bibinfo{author}{\bibfnamefont{C.}~\bibnamefont{Hanhart}},
  \bibinfo{author}{\bibfnamefont{J.~R.} \bibnamefont{Pelaez}},
  \bibnamefont{and} \bibinfo{author}{\bibfnamefont{G.}~\bibnamefont{Rios}},
  \bibinfo{journal}{Phys. Rev. Lett.} \textbf{\bibinfo{volume}{100}},
  \bibinfo{pages}{152001} (\bibinfo{year}{2008}), \eprint{0801.2871}.

\bibitem[{\citenamefont{Pelaez and Rios}(2010)}]{Pelaez:2010fj}
\bibinfo{author}{\bibfnamefont{J.~R.} \bibnamefont{Pelaez}} \bibnamefont{and}
  \bibinfo{author}{\bibfnamefont{G.}~\bibnamefont{Rios}},
  \bibinfo{journal}{Phys. Rev.} \textbf{\bibinfo{volume}{D82}},
  \bibinfo{pages}{114002} (\bibinfo{year}{2010}), \eprint{1010.6008}.

\bibitem[{\citenamefont{Liu et~al.}(2013)\citenamefont{Liu, Orginos, Guo,
  Hanhart, and Meissner}}]{Liu:2012zya}
\bibinfo{author}{\bibfnamefont{L.}~\bibnamefont{Liu}},
  \bibinfo{author}{\bibfnamefont{K.}~\bibnamefont{Orginos}},
  \bibinfo{author}{\bibfnamefont{F.-K.} \bibnamefont{Guo}},
  \bibinfo{author}{\bibfnamefont{C.}~\bibnamefont{Hanhart}}, \bibnamefont{and}
  \bibinfo{author}{\bibfnamefont{U.-G.} \bibnamefont{Meissner}},
  \bibinfo{journal}{Phys. Rev.} \textbf{\bibinfo{volume}{D87}},
  \bibinfo{pages}{014508} (\bibinfo{year}{2013}), \eprint{1208.4535}.

\bibitem[{\citenamefont{Rummukainen and Gottlieb}(1995)}]{Rummukainen:1995vs}
\bibinfo{author}{\bibfnamefont{K.}~\bibnamefont{Rummukainen}} \bibnamefont{and}
  \bibinfo{author}{\bibfnamefont{S.~A.} \bibnamefont{Gottlieb}},
  \bibinfo{journal}{Nucl. Phys.} \textbf{\bibinfo{volume}{B450}},
  \bibinfo{pages}{397} (\bibinfo{year}{1995}), \eprint{hep-lat/9503028}.

\bibitem[{\citenamefont{Kim et~al.}(2005)\citenamefont{Kim, Sachrajda, and
  Sharpe}}]{Kim:2005gf}
\bibinfo{author}{\bibfnamefont{C.~h.} \bibnamefont{Kim}},
  \bibinfo{author}{\bibfnamefont{C.~T.} \bibnamefont{Sachrajda}},
  \bibnamefont{and} \bibinfo{author}{\bibfnamefont{S.~R.}
  \bibnamefont{Sharpe}}, \bibinfo{journal}{Nucl. Phys.}
  \textbf{\bibinfo{volume}{B727}}, \bibinfo{pages}{218} (\bibinfo{year}{2005}),
  \eprint{hep-lat/0507006}.

\bibitem[{\citenamefont{Christ et~al.}(2005)\citenamefont{Christ, Kim, and
  Yamazaki}}]{Christ:2005gi}
\bibinfo{author}{\bibfnamefont{N.~H.} \bibnamefont{Christ}},
  \bibinfo{author}{\bibfnamefont{C.}~\bibnamefont{Kim}}, \bibnamefont{and}
  \bibinfo{author}{\bibfnamefont{T.}~\bibnamefont{Yamazaki}},
  \bibinfo{journal}{Phys. Rev.} \textbf{\bibinfo{volume}{D72}},
  \bibinfo{pages}{114506} (\bibinfo{year}{2005}), \eprint{hep-lat/0507009}.

\bibitem[{\citenamefont{Bedaque et~al.}(2006)\citenamefont{Bedaque, Sato, and
  Walker-Loud}}]{Bedaque:2006yi}
\bibinfo{author}{\bibfnamefont{P.~F.} \bibnamefont{Bedaque}},
  \bibinfo{author}{\bibfnamefont{I.}~\bibnamefont{Sato}}, \bibnamefont{and}
  \bibinfo{author}{\bibfnamefont{A.}~\bibnamefont{Walker-Loud}},
  \bibinfo{journal}{Phys. Rev.} \textbf{\bibinfo{volume}{D73}},
  \bibinfo{pages}{074501} (\bibinfo{year}{2006}), \eprint{hep-lat/0601033}.

\bibitem[{\citenamefont{Weinberg}(1966)}]{Weinberg:1966kf}
\bibinfo{author}{\bibfnamefont{S.}~\bibnamefont{Weinberg}},
  \bibinfo{journal}{Phys. Rev. Lett.} \textbf{\bibinfo{volume}{17}},
  \bibinfo{pages}{616} (\bibinfo{year}{1966}).

\bibitem[{\citenamefont{Colangelo et~al.}(2001)\citenamefont{Colangelo, Gasser,
  and Leutwyler}}]{Colangelo:2001df}
\bibinfo{author}{\bibfnamefont{G.}~\bibnamefont{Colangelo}},
  \bibinfo{author}{\bibfnamefont{J.}~\bibnamefont{Gasser}}, \bibnamefont{and}
  \bibinfo{author}{\bibfnamefont{H.}~\bibnamefont{Leutwyler}},
  \bibinfo{journal}{Nucl. Phys.} \textbf{\bibinfo{volume}{B603}},
  \bibinfo{pages}{125} (\bibinfo{year}{2001}), \eprint{hep-ph/0103088}.

\bibitem[{\citenamefont{Ecker et~al.}(1989)\citenamefont{Ecker, Gasser, Pich,
  and de~Rafael}}]{Ecker:1988te}
\bibinfo{author}{\bibfnamefont{G.}~\bibnamefont{Ecker}},
  \bibinfo{author}{\bibfnamefont{J.}~\bibnamefont{Gasser}},
  \bibinfo{author}{\bibfnamefont{A.}~\bibnamefont{Pich}}, \bibnamefont{and}
  \bibinfo{author}{\bibfnamefont{E.}~\bibnamefont{de~Rafael}},
  \bibinfo{journal}{Nucl. Phys.} \textbf{\bibinfo{volume}{B321}},
  \bibinfo{pages}{311} (\bibinfo{year}{1989}).

\bibitem[{\citenamefont{Gasser and Leutwyler}(1985)}]{Gasser:1984gg}
\bibinfo{author}{\bibfnamefont{J.}~\bibnamefont{Gasser}} \bibnamefont{and}
  \bibinfo{author}{\bibfnamefont{H.}~\bibnamefont{Leutwyler}},
  \bibinfo{journal}{Nucl. Phys.} \textbf{\bibinfo{volume}{B250}},
  \bibinfo{pages}{465} (\bibinfo{year}{1985}).

\bibitem[{\citenamefont{Gasser and Leutwyler}(1984)}]{Gasser:1983yg}
\bibinfo{author}{\bibfnamefont{J.}~\bibnamefont{Gasser}} \bibnamefont{and}
  \bibinfo{author}{\bibfnamefont{H.}~\bibnamefont{Leutwyler}},
  \bibinfo{journal}{Annals Phys.} \textbf{\bibinfo{volume}{158}},
  \bibinfo{pages}{142} (\bibinfo{year}{1984}).

\bibitem[{\citenamefont{Gasser and Leutwyler}(1983)}]{Gasser:1983kx}
\bibinfo{author}{\bibfnamefont{J.}~\bibnamefont{Gasser}} \bibnamefont{and}
  \bibinfo{author}{\bibfnamefont{H.}~\bibnamefont{Leutwyler}},
  \bibinfo{journal}{Phys. Lett.} \textbf{\bibinfo{volume}{B125}},
  \bibinfo{pages}{325} (\bibinfo{year}{1983}).

\bibitem[{\citenamefont{Mastropas and Richards}(2014)}]{Mastropas:2014fsa}
\bibinfo{author}{\bibfnamefont{E.~V.} \bibnamefont{Mastropas}}
  \bibnamefont{and} \bibinfo{author}{\bibfnamefont{D.~G.}
  \bibnamefont{Richards}} (\bibinfo{collaboration}{Hadron Spectrum}),
  \bibinfo{journal}{Phys. Rev.} \textbf{\bibinfo{volume}{D90}},
  \bibinfo{pages}{014511} (\bibinfo{year}{2014}), \eprint{1403.5575}.

\bibitem[{\citenamefont{Dürr}(2015)}]{Durr:2014oba}
\bibinfo{author}{\bibfnamefont{S.}~\bibnamefont{Dürr}}, \bibinfo{journal}{PoS}
  \textbf{\bibinfo{volume}{LATTICE2014}}, \bibinfo{pages}{006}
  (\bibinfo{year}{2015}), \eprint{1412.6434}.

\bibitem[{\citenamefont{Gomez~Nicola et~al.}(2008)\citenamefont{Gomez~Nicola,
  Pelaez, and Rios}}]{GomezNicola:2007qj}
\bibinfo{author}{\bibfnamefont{A.}~\bibnamefont{Gomez~Nicola}},
  \bibinfo{author}{\bibfnamefont{J.~R.} \bibnamefont{Pelaez}},
  \bibnamefont{and} \bibinfo{author}{\bibfnamefont{G.}~\bibnamefont{Rios}},
  \bibinfo{journal}{Phys. Rev.} \textbf{\bibinfo{volume}{D77}},
  \bibinfo{pages}{056006} (\bibinfo{year}{2008}), \eprint{0712.2763}.

\bibitem[{\citenamefont{Ananthanarayan
  et~al.}(2001)\citenamefont{Ananthanarayan, Colangelo, Gasser, and
  Leutwyler}}]{Ananthanarayan:2000ht}
\bibinfo{author}{\bibfnamefont{B.}~\bibnamefont{Ananthanarayan}},
  \bibinfo{author}{\bibfnamefont{G.}~\bibnamefont{Colangelo}},
  \bibinfo{author}{\bibfnamefont{J.}~\bibnamefont{Gasser}}, \bibnamefont{and}
  \bibinfo{author}{\bibfnamefont{H.}~\bibnamefont{Leutwyler}},
  \bibinfo{journal}{Phys. Rept.} \textbf{\bibinfo{volume}{353}},
  \bibinfo{pages}{207} (\bibinfo{year}{2001}), \eprint{hep-ph/0005297}.

\bibitem[{\citenamefont{Garcia-Martin
  et~al.}(2011{\natexlab{a}})\citenamefont{Garcia-Martin, Kaminski, Pelaez,
  Ruiz~de Elvira, and Yndurain}}]{GarciaMartin:2011cn}
\bibinfo{author}{\bibfnamefont{R.}~\bibnamefont{Garcia-Martin}},
  \bibinfo{author}{\bibfnamefont{R.}~\bibnamefont{Kaminski}},
  \bibinfo{author}{\bibfnamefont{J.~R.} \bibnamefont{Pelaez}},
  \bibinfo{author}{\bibfnamefont{J.}~\bibnamefont{Ruiz~de Elvira}},
  \bibnamefont{and} \bibinfo{author}{\bibfnamefont{F.~J.}
  \bibnamefont{Yndurain}}, \bibinfo{journal}{Phys. Rev.}
  \textbf{\bibinfo{volume}{D83}}, \bibinfo{pages}{074004}
  (\bibinfo{year}{2011}{\natexlab{a}}), \eprint{1102.2183}.

\bibitem[{\citenamefont{Danilkin et~al.}(2011)\citenamefont{Danilkin, Gil, and
  Lutz}}]{Danilkin:2011fz}
\bibinfo{author}{\bibfnamefont{I.~V.} \bibnamefont{Danilkin}},
  \bibinfo{author}{\bibfnamefont{L.~I.~R.} \bibnamefont{Gil}},
  \bibnamefont{and} \bibinfo{author}{\bibfnamefont{M.~F.~M.}
  \bibnamefont{Lutz}}, \bibinfo{journal}{Phys. Lett.}
  \textbf{\bibinfo{volume}{B703}}, \bibinfo{pages}{504} (\bibinfo{year}{2011}),
  \eprint{1106.2230}.

\bibitem[{\citenamefont{Lin et~al.}(2009)}]{Lin:2008pr}
\bibinfo{author}{\bibfnamefont{H.-W.} \bibnamefont{Lin}} \bibnamefont{et~al.}
  (\bibinfo{collaboration}{Hadron Spectrum}), \bibinfo{journal}{Phys. Rev.}
  \textbf{\bibinfo{volume}{D79}}, \bibinfo{pages}{034502}
  (\bibinfo{year}{2009}), \eprint{0810.3588}.

\bibitem[{\citenamefont{Protopopescu et~al.}(1973)\citenamefont{Protopopescu,
  Alston-Garnjost, Barbaro-Galtieri, Flatte, Friedman, Lasinski, Lynch, Rabin,
  and Solmitz}}]{Protopopescu:1973sh}
\bibinfo{author}{\bibfnamefont{S.~D.} \bibnamefont{Protopopescu}},
  \bibinfo{author}{\bibfnamefont{M.}~\bibnamefont{Alston-Garnjost}},
  \bibinfo{author}{\bibfnamefont{A.}~\bibnamefont{Barbaro-Galtieri}},
  \bibinfo{author}{\bibfnamefont{S.~M.} \bibnamefont{Flatte}},
  \bibinfo{author}{\bibfnamefont{J.~H.} \bibnamefont{Friedman}},
  \bibinfo{author}{\bibfnamefont{T.~A.} \bibnamefont{Lasinski}},
  \bibinfo{author}{\bibfnamefont{G.~R.} \bibnamefont{Lynch}},
  \bibinfo{author}{\bibfnamefont{M.~S.} \bibnamefont{Rabin}}, \bibnamefont{and}
  \bibinfo{author}{\bibfnamefont{F.~T.} \bibnamefont{Solmitz}},
  \bibinfo{journal}{Phys. Rev.} \textbf{\bibinfo{volume}{D7}},
  \bibinfo{pages}{1279} (\bibinfo{year}{1973}).

\bibitem[{\citenamefont{Estabrooks and Martin}(1974)}]{Estabrooks:1974vu}
\bibinfo{author}{\bibfnamefont{P.}~\bibnamefont{Estabrooks}} \bibnamefont{and}
  \bibinfo{author}{\bibfnamefont{A.~D.} \bibnamefont{Martin}},
  \bibinfo{journal}{Nucl. Phys.} \textbf{\bibinfo{volume}{B79}},
  \bibinfo{pages}{301} (\bibinfo{year}{1974}).

\bibitem[{\citenamefont{Masjuan et~al.}(2014)\citenamefont{Masjuan, Ruiz~de
  Elvira, and Sanz-Cillero}}]{Masjuan:2014psa}
\bibinfo{author}{\bibfnamefont{P.}~\bibnamefont{Masjuan}},
  \bibinfo{author}{\bibfnamefont{J.}~\bibnamefont{Ruiz~de Elvira}},
  \bibnamefont{and} \bibinfo{author}{\bibfnamefont{J.~J.}
  \bibnamefont{Sanz-Cillero}}, \bibinfo{journal}{Phys. Rev.}
  \textbf{\bibinfo{volume}{D90}}, \bibinfo{pages}{097901}
  (\bibinfo{year}{2014}), \eprint{1410.2397}.

\bibitem[{\citenamefont{Zhou et~al.}(2005)\citenamefont{Zhou, Qin, Zhang, Xiao,
  Zheng, and Wu}}]{Zhou:2004ms}
\bibinfo{author}{\bibfnamefont{Z.~Y.} \bibnamefont{Zhou}},
  \bibinfo{author}{\bibfnamefont{G.~Y.} \bibnamefont{Qin}},
  \bibinfo{author}{\bibfnamefont{P.}~\bibnamefont{Zhang}},
  \bibinfo{author}{\bibfnamefont{Z.}~\bibnamefont{Xiao}},
  \bibinfo{author}{\bibfnamefont{H.~Q.} \bibnamefont{Zheng}}, \bibnamefont{and}
  \bibinfo{author}{\bibfnamefont{N.}~\bibnamefont{Wu}}, \bibinfo{journal}{JHEP}
  \textbf{\bibinfo{volume}{02}}, \bibinfo{pages}{043} (\bibinfo{year}{2005}),
  \eprint{hep-ph/0406271}.

\bibitem[{\citenamefont{Garcia-Martin
  et~al.}(2011{\natexlab{b}})\citenamefont{Garcia-Martin, Kaminski, Pelaez, and
  Ruiz~de Elvira}}]{GarciaMartin:2011jx}
\bibinfo{author}{\bibfnamefont{R.}~\bibnamefont{Garcia-Martin}},
  \bibinfo{author}{\bibfnamefont{R.}~\bibnamefont{Kaminski}},
  \bibinfo{author}{\bibfnamefont{J.~R.} \bibnamefont{Pelaez}},
  \bibnamefont{and} \bibinfo{author}{\bibfnamefont{J.}~\bibnamefont{Ruiz~de
  Elvira}}, \bibinfo{journal}{Phys. Rev. Lett.} \textbf{\bibinfo{volume}{107}},
  \bibinfo{pages}{072001} (\bibinfo{year}{2011}{\natexlab{b}}),
  \eprint{1107.1635}.

\bibitem[{\citenamefont{Masjuan and Sanz-Cillero}(2013)}]{Masjuan:2013jha}
\bibinfo{author}{\bibfnamefont{P.}~\bibnamefont{Masjuan}} \bibnamefont{and}
  \bibinfo{author}{\bibfnamefont{J.~J.} \bibnamefont{Sanz-Cillero}},
  \bibinfo{journal}{Eur. Phys. J.} \textbf{\bibinfo{volume}{C73}},
  \bibinfo{pages}{2594} (\bibinfo{year}{2013}), \eprint{1306.6308}.

\bibitem[{\citenamefont{Roy}(1971)}]{Roy:1971tc}
\bibinfo{author}{\bibfnamefont{S.~M.} \bibnamefont{Roy}},
  \bibinfo{journal}{Phys. Lett.} \textbf{\bibinfo{volume}{B36}},
  \bibinfo{pages}{353} (\bibinfo{year}{1971}).

\bibitem[{\citenamefont{Dudek et~al.}(2011{\natexlab{b}})\citenamefont{Dudek,
  Edwards, Peardon, Richards, and Thomas}}]{Dudek:2010ew}
\bibinfo{author}{\bibfnamefont{J.~J.} \bibnamefont{Dudek}},
  \bibinfo{author}{\bibfnamefont{R.~G.} \bibnamefont{Edwards}},
  \bibinfo{author}{\bibfnamefont{M.~J.} \bibnamefont{Peardon}},
  \bibinfo{author}{\bibfnamefont{D.~G.} \bibnamefont{Richards}},
  \bibnamefont{and} \bibinfo{author}{\bibfnamefont{C.~E.}
  \bibnamefont{Thomas}}, \bibinfo{journal}{Phys. Rev.}
  \textbf{\bibinfo{volume}{D83}}, \bibinfo{pages}{071504}
  (\bibinfo{year}{2011}{\natexlab{b}}), \eprint{1011.6352}.

\bibitem[{\citenamefont{Dudek et~al.}(2012)\citenamefont{Dudek, Edwards, and
  Thomas}}]{Dudek:2012gj}
\bibinfo{author}{\bibfnamefont{J.~J.} \bibnamefont{Dudek}},
  \bibinfo{author}{\bibfnamefont{R.~G.} \bibnamefont{Edwards}},
  \bibnamefont{and} \bibinfo{author}{\bibfnamefont{C.~E.}
  \bibnamefont{Thomas}}, \bibinfo{journal}{Phys. Rev.}
  \textbf{\bibinfo{volume}{D86}}, \bibinfo{pages}{034031}
  (\bibinfo{year}{2012}), \eprint{1203.6041}.

\bibitem[{\citenamefont{Martínez~Torres
  et~al.}(2015)\citenamefont{Martínez~Torres, Oset, Prelovsek, and
  Ramos}}]{Torres:2014vna}
\bibinfo{author}{\bibfnamefont{A.}~\bibnamefont{Martínez~Torres}},
  \bibinfo{author}{\bibfnamefont{E.}~\bibnamefont{Oset}},
  \bibinfo{author}{\bibfnamefont{S.}~\bibnamefont{Prelovsek}},
  \bibnamefont{and} \bibinfo{author}{\bibfnamefont{A.}~\bibnamefont{Ramos}},
  \bibinfo{journal}{JHEP} \textbf{\bibinfo{volume}{05}}, \bibinfo{pages}{153}
  (\bibinfo{year}{2015}), \eprint{1412.1706}.

\bibitem[{\citenamefont{Briceno
  et~al.}(2015{\natexlab{b}})\citenamefont{Briceno, Dudek, Edwards, Shultz,
  Thomas, and Wilson}}]{Briceno:2015dca}
\bibinfo{author}{\bibfnamefont{R.~A.} \bibnamefont{Briceno}},
  \bibinfo{author}{\bibfnamefont{J.~J.} \bibnamefont{Dudek}},
  \bibinfo{author}{\bibfnamefont{R.~G.} \bibnamefont{Edwards}},
  \bibinfo{author}{\bibfnamefont{C.~J.} \bibnamefont{Shultz}},
  \bibinfo{author}{\bibfnamefont{C.~E.} \bibnamefont{Thomas}},
  \bibnamefont{and} \bibinfo{author}{\bibfnamefont{D.~J.}
  \bibnamefont{Wilson}}, \bibinfo{journal}{Phys. Rev. Lett.}
  \textbf{\bibinfo{volume}{115}}, \bibinfo{pages}{242001}
  (\bibinfo{year}{2015}{\natexlab{b}}), \eprint{1507.06622}.

\end{thebibliography}

\end{document}